\documentclass[12pt]{iopart}
\expandafter\let\csname equation*\endcsname\relax
\expandafter\let\csname endequation*\endcsname\relax
\usepackage{amsmath}

\usepackage{braket} % bra ket notation
\usepackage{graphicx}
\usepackage{algorithm}
\usepackage[noend]{algpseudocode}
\usepackage[colorlinks=true,citecolor=blue,linkcolor=black]{hyperref}
\usepackage{color}
\usepackage[dvipsnames]{xcolor}
\usepackage[normalem]{ulem}

\usepackage{caption}
\usepackage{subcaption}

\newcommand{\tab}[1]{\makebox[.04\linewidth][l]{#1}\ignorespaces}%
\newcommand{\tabb}[1]{\makebox[.08\linewidth][l]{#1}\ignorespaces}%

\usepackage{soul}

\usepackage{tikz,xcolor,hyperref}
\definecolor{lime}{HTML}{A6CE39}
\DeclareRobustCommand{\orcidicon}{
    \hspace{-3mm}
	\begin{tikzpicture}
	\draw[lime, fill=lime] (0,0) 
	circle [radius=0.16] 
	node[white] {{\fontfamily{qag}\selectfont \tiny ID}};
	\draw[white, fill=white] (-0.0625,0.095) 
	circle [radius=0.007];
	\end{tikzpicture}
	\hspace{-4mm}
}

\foreach \x in {A, ..., Z}{%
	\expandafter\xdef\csname orcid\x\endcsname{\noexpand\href{https://orcid.org/\csname orcidauthor\x\endcsname}{\noexpand\orcidicon}}
}

\begin{document}

\title[]{Deep learning based quantum vortex detection in atomic Bose-Einstein condensates}

\author{Friederike Metz\orcidA{}, Juan Polo\orcidB{}, Natalya Weber\orcidC{} and Thomas Busch\orcidD{}}
\address{Quantum Systems Unit, Okinawa Institute of Science and Technology Graduate University, 1919-1 Tancha, Onna, Okinawa 904-0495, Japan}
\ead{\mailto{friederike.metz@oist.jp}}
\vspace{10pt}
%\begin{indented}
%\item[]August 2017
%\end{indented}

\begin{abstract}
Quantum vortices naturally emerge in rotating Bose-Einstein condensates (BECs) and, similarly to their classical counterparts, allow the study of a range of interesting out-of-equilibrium phenomena like turbulence and chaos. However, the study of such phenomena requires to determine the precise location of each vortex within a BEC, which becomes challenging when either only the condensate density is available or sources of noise are present, as is typically the case in experimental settings. Here, we introduce a machine learning based vortex detector motivated by state-of-the-art object detection methods that can accurately locate vortices in simulated BEC density images. Our model allows for robust and real-time detection in noisy and non-equilibrium configurations. Furthermore, the network can distinguish between vortices and anti-vortices if the condensate phase profile is also available. We anticipate that our vortex detector will be advantageous both for experimental and theoretical studies of the static and dynamical properties of vortex configurations in BECs.
\end{abstract}

\vspace{2pc}
\noindent{\it Keywords}: machine learning, object detection, convolutional neural network, vortices, Bose-Einstein condensate, non-equilibrium dynamics, Gross–Pitaevskii equation
%\maketitle

\section{Introduction}

Non-equilibrium behaviour of classical and quantum systems is ubiquitous in nature and includes interesting and complex phenomena such as turbulence and chaos, which are still only partially understood \cite{dorfman99,spicka19}. Bose-Einstein condensates (BECs) provide a particularly versatile platform for studying and simulating general features of non-equilibrium dynamics, due to the high level of control over the experimental systems \cite{Dalfovo99,Fetter01}. In particular, rapidly rotating BECs can support quantum vortices, which are considered a key component of superfluid turbulence \cite{White14}. Unlike their classical counterparts quantum vortices are restricted to quantized circulation due to the condition that the wave function has to be single valued at all points. This leads to a well-defined velocity profile that is given by the gradient of the phase \cite{Parker08}. Numerous experiments have generated vortices in BECs \cite{Chevy08} and observed the formation of vortex-antivortex pairs \cite{Inouye01}, vortex rings \cite{Anderson01}, and vortex lattices \cite{Abo-Shaeer01}. Furthermore, in-situ density imaging of vortex cores has opened the door to the analysis of their real-time dynamics \cite{Freilich10,Wilson15,Haljan01} and thus, the experimental study of chaos, turbulence, and other out-of-equilibrium dynamics \cite{Navarro13,Serafini17, Kwon14,Neely13,Reeves20}. For example, recent results include the detection of persistent vortex clusters emerging from the turbulent flow of high-energy vortex configurations \cite{Johnstone19,Gauthier19}, the experimental realization of the quantum analogue of the Kármán vortex street \cite{Kwon16}, and the observation of vortex-antivortex pairing in a turbulent BEC \cite{Seo17}.

However, the study of quantized vortices and specifically their dynamics requires to first infer their precise location within a BEC \cite{Rakonjac16}. For ground states the task of detecting vortices is straightforward, since they are arranged in a clear pattern with pronounced density minima at their core centers \cite{Abo-Shaeer01,Aftalion01} and therefore can be easily spotted by eye or via automated processes. On the other hand, in non-equilibrium configurations vortices are located at random positions following no distinct order. Furthermore, local density minima not corresponding to vortices can emerge as a consequence of phononic excitations, making the detection of vortices considerably more difficult \cite{Ortega19}. In numerical simulations that model the dynamics of BECs one usually has access to the full condensate wave function and hence also to its phase. The phase profile provides a clear indication of the existence of a vortex through a phase winding of $2\pi$ around the position of a vortex core. Therefore, vortex detection algorithms for non-equilibrium configurations mainly rely on the BEC phase profile to distinguish vortices from other defects \cite{Ortega19,Groszek20}. However, in experiments the phase profile and thus the information encoded therein is not easily accessible. Moreover, non zero temperatures and the presence of noise pose an additional challenge for accurately detecting vortices and hence require the development of more elaborate methods. 

In this paper we show that a machine learning based vortex detector can reliably and accurately locate vortices within out-of-equilibrium BEC density images. It can distinguish vortices from other local density minima even in situations that are difficult for the human eye. In contrast to conventional vortex detection algorithms, such as blob detection, the neural network does not require hard-coded features or fine tuning of parameters \cite{Rakonjac16, Groszek20}. In addition, the model is robust, i.e.~it performs well on simulated data with experimentally relevant sources of noise and generalizes to configurations it has not been trained on, which would not be possible with more traditional object detection methods like template matching \cite{Brunelli09}. Hence, we anticipate that our vortex detector can be broadly employed in experimental studies of non-equilibrium vortex configurations where only the BEC density is accessible. On the other hand, in numerical simulations of the BEC the phase profile is available and can be provided to the neural network as additional information. In this case the model is also able to accurately classify the circulation direction of each vortex.

In recent years machine learning techniques have become a widely adopted tool in the field of quantum physics \cite{Dunjko18,Carrasquilla20}. Specifically in the area of BECs, machine learning methods have been used to optimize the cooling process for the atomic gas \cite{Barker20}, learn the Kosterlitz-Thouless transition \cite{Beach18}, and devise control schemes for the creation of quantum vortices \cite{Saito20}. On the other hand, deep learning based object detection has celebrated remarkable successes in the field of classical computer vision, achieving state-of-the-art results in areas like face, vehicle, and medical image detection \cite{Xiao20, Liu20}. Hence, neural network based object detection promises to be a powerful tool for the physical sciences as well and has already been successfully employed in a few cases \cite{Minor20,Usman20}, as for instance to detect and identify characteristics of atomic clouds \cite{Hofer20} or to locate dark solitons in a BEC \cite{guo2021}. Finally, let us note that deep learning approaches have also been applied to the detection of vortices in classical fluids such as locating rotor blade tip vortices \cite{Luo20} or eddies in ocean currents \cite{Bai19}. Motivated by these recent successes, in this work we employ a convolutional neural network (CNN) ansatz for the task of vortex detection which can achieve high accuracies on our test data and is therefore very well suited for the problem of locating vortices in BECs.

The manuscript is organized as follows:~We first present the theoretical model used to simulate BECs and describe how vortices emerge in Section \ref{sec:model}. Section \ref{sec:cnn} introduces the machine learning based vortex detector. The results of training the model on BEC density images alone are discussed in Section \ref{sec:density} and the case of training with density and phase snapshots is presented in Section \ref{sec:phase}. In the appendices we provide further details and discussion on the training data, the network architecture, the evaluation metrics, the features learned by the CNN, and the ability of the network to generalize to different trapping geometries and different levels of noise.

\section{Physical system}\label{sec:model}

We consider a dilute and weakly-interacting Bose-Einstein condensate rotating around the $z$-axis with rotational frequency $\Omega$. At zero temperature and assuming a tight harmonic confinement in the $z$ direction such that the transverse dynamics is frozen out, i.e.~$\omega_z\gg \omega_\perp$, we can describe the dynamics of the Bose gas in the co-rotating frame by means of the two-dimensional mean field Gross–Pitaevskii equation (GPE) of the form \cite{Salasnich02, Petrov00}
\begin{equation}\label{eq:01}
	i \hbar \frac{\partial}{\partial t} \Psi=\left(-\frac{\hbar^{2}}{2 m} \nabla^{2}+\frac{1}{2} m \omega_{\perp}^{2} r^{2}+g|\Psi|^{2}-\Omega L_{z}\right) \Psi ,
\end{equation}
with $\Psi$ being the condensate wave function, $\omega_{\perp}$ the frequency of the harmonic trap, $L_{z}=x p_{y}-y p_{x}$ the angular momentum operator, and $r=\sqrt{x^2 + y^2}$ the radial distance. The effective two-dimensional interaction strength is given by $g = g_{3D}/(\sqrt{2\pi}a_z)$  with $a_z=\sqrt{\hbar/m\omega_z}$ and $g_{3D}=4\pi\hbar^2a_s/m$ being the harmonic oscillator length scale of the transverse tight confinement and the three-dimensional interatomic interaction strength respectively. Here $a_s$ is the s-wave scattering length. Note that equation \eqref{eq:01} is analogous to the more general nonlinear Schr\"odinger equation which can describe a variety of different systems \cite{liu19}. From here onward we use harmonic oscillator units by setting $\hbar = \omega_{\perp} = m = 1$ and choose interaction strengths $g\in [50,600]$ as well as rotation frequencies $\Omega\in [0.65, 0.95]$ which correspond to experimentally accessible parameter regimes.

\begin{figure}[t]
	\centering
	\hspace*{-33mm}
	\begin{subfigure}[b]{0.25\textwidth}
		\centering
		\includegraphics[width=\textwidth]{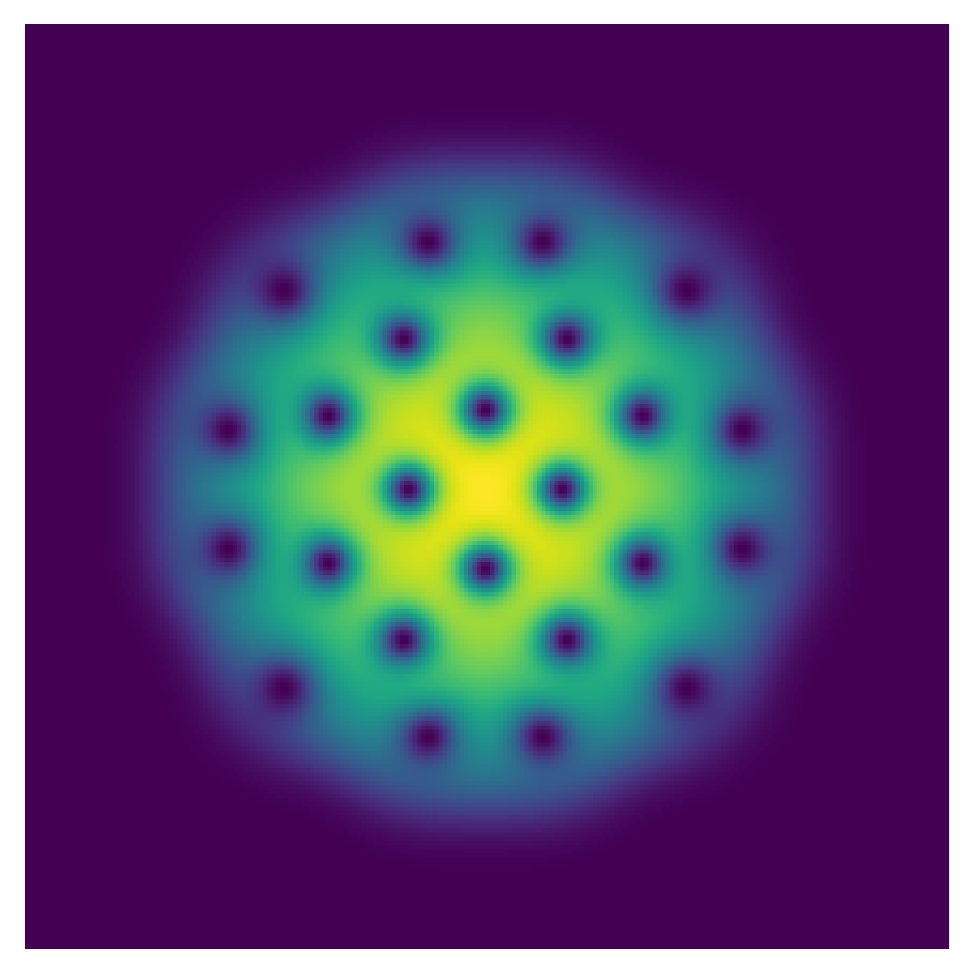}
		\caption{}
	\end{subfigure}
	\begin{subfigure}[b]{0.25\textwidth}
		\centering
		\includegraphics[width=\textwidth]{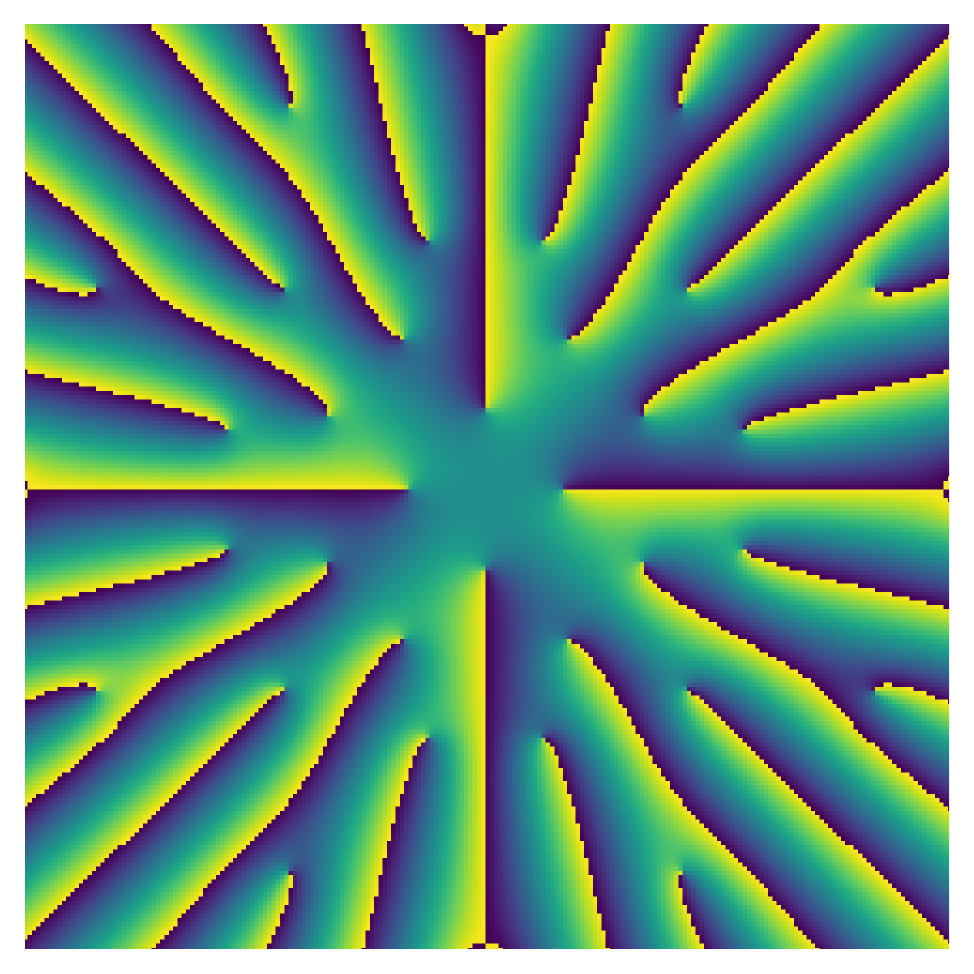}
		\caption{}
	\end{subfigure}
	\begin{subfigure}[b]{0.25\textwidth}
		\centering
		\includegraphics[width=\textwidth]{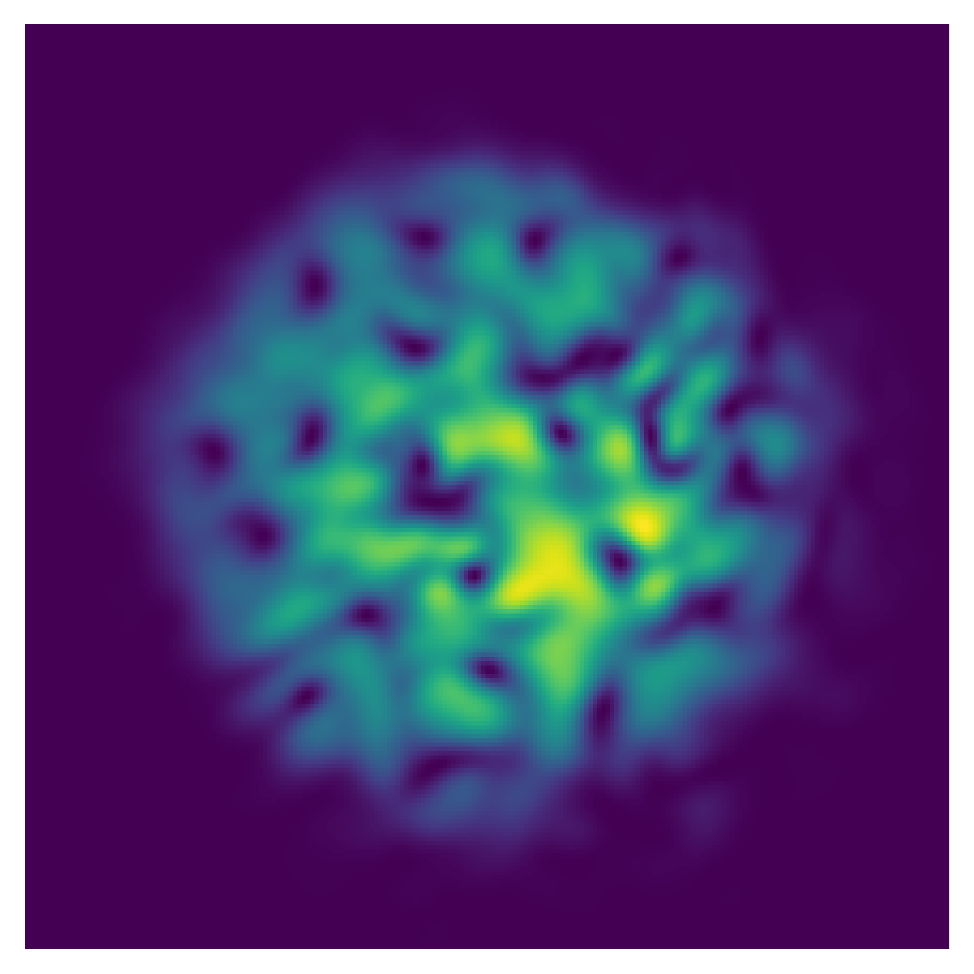}
		\caption{}
	\end{subfigure}
	\begin{subfigure}[b]{0.25\textwidth}
		\centering
		\includegraphics[width=\textwidth]{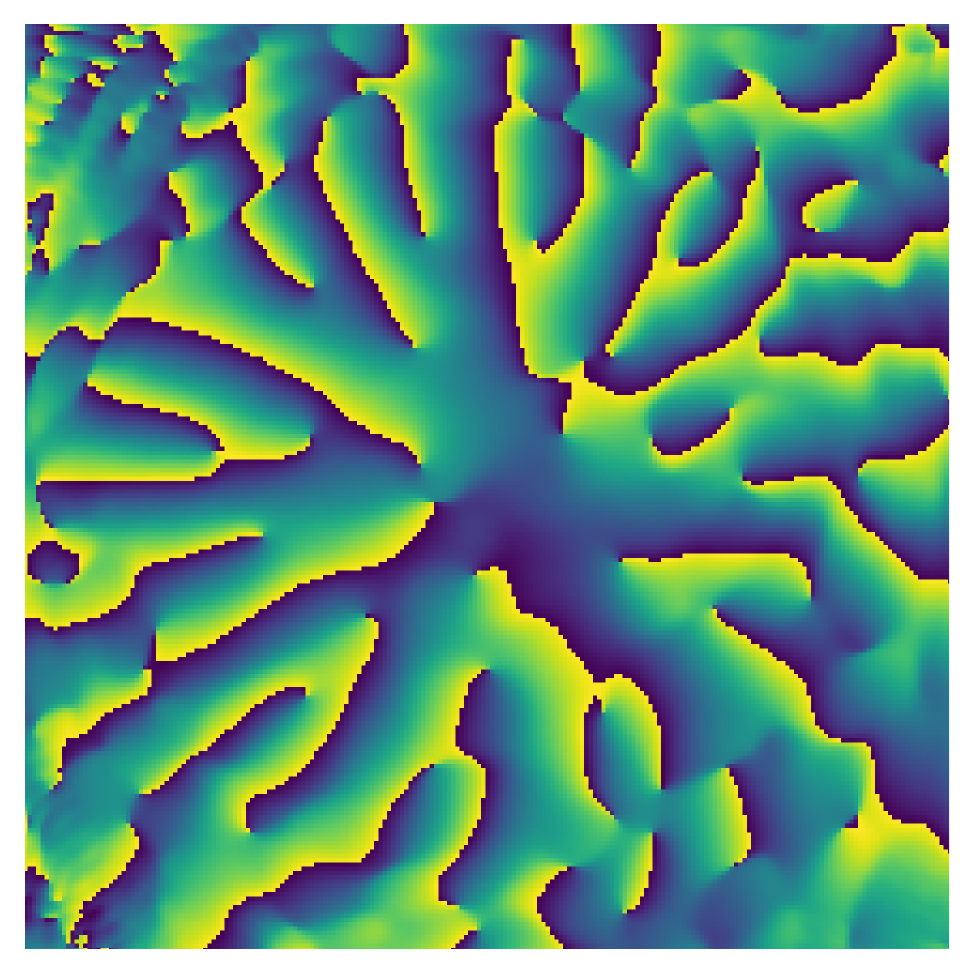}
		\caption{}
	\end{subfigure}
	\hspace*{-33mm}
	\caption{Examples of BEC density and phase profiles for the stationary ground state (a),(b) and for a non-equilibrium configuration (c),(d). The ground state is computed via imaginary time evolution with the GPE using an interaction strength $g=452$ and a rotation frequency $\Omega=0.816$. Phase imprinting of additional vortices and a subsequent real time evolution gives rise to the out-of-equilibrium configuration.}
	\label{fig:01}
\end{figure}

Above a critical rotation frequency $\Omega_c$, the ground state of Eq.~\eqref{eq:01} possesses  vortices   \cite{Lundh97, Chevy00, Madison00}. For large rotation frequencies these vortices  arrange themselves in a triangular lattice geometry \cite{Abo-Shaeer01}, while for smaller frequencies different configurations can arise \cite{Aftalion01}. As an example, figure \ref{fig:01}(a)-(b) shows the numerically obtained density distribution $|\Psi(\mathbf{r})|^2$ and phase profile of the ground state wave function when $g=452$ and $\Omega=0.816$. The vortices are clearly defined through a density dip at their cores and through the characteristic $2\pi$ phase winding in the phase. Note that the detailed structure of the vortex core depends on the trapping potential \cite{Fetter01, Parker08}: in a  homogeneous BEC, the width of a vortex core is fixed by the balance between the kinetic and interaction energy, with a typical core size given by the healing length $\xi=\sqrt{8\pi n a_s}$, where $n$ corresponds to the density. In trap systems, the size of the vortex core depends also on the local chemical potential, which gives rise to slightly larger sizes in low density regions. In addition, vortices surrounded by very low densities at the outer part of the BEC will not contribute to the rotational energy of the system and are therefore irrelevant from a physical point of view \cite{Tsubota02}.

The vortices carried by the ground state all rotate in the same direction, i.e.~have a winding number with the same sign, which is determined by the rotation frequency $\Omega$. Situations where vortices of different rotation directions co-exist can be created for instance by forcing the superfluid to flow around an obstacle potential \cite{Sasaki10,Neely10} or through the process of phase imprinting \cite{Riordan16,James19,Leanhardt02,Dobrek99,Andersen06}. In the latter case, a single vortex centered at $ (x_0,y_0) $ is generated by applying a phase mask $\phi_{\mathrm{IMP}}(\mathbf{r})=\arctan \left(y-y_{0}, x-x_{0}\right)$ with a $2\pi$ phase winding in the desired direction. The time-evolution of configurations with multiple vortices of unequal rotation direction features interesting out-of-equilibrium processes such as vortex - antivortex annihilation and the emergence of other low energy excitations. Furthermore, it has been shown that a three and four vortex-system with one counter-rotating vortex can already lead to chaotic dynamics \cite{Navarro13,James19,Koukouloyannis14} and that large vortex systems can give rise to quantum turbulence \cite{White14,Johnstone19,Wilson13,White10}. Figure \ref{fig:01}(c)-(d) displays a density and phase profile snapshot during a representative time evolution after phase imprinting additional anti-vortices. While the vortex cores are still clearly visible in the image of the condensate phase, it is more challenging to pinpoint their exact location in the density snapshot.

\section{Machine learning model}\label{sec:cnn}

In the following we introduce our neural network based vortex detector which is motivated by state-of-the-art object detectors such as YOLO and Objects as Points \cite{Redmon16, Zhou19}. The general task of object detection is to locate each object in an image, draw the corresponding bounding boxes, and associate them to a specific class. Here, we are only interested in detecting vortices and therefore our problem reduces to that of binary classification. In Section \ref{sec:phase} we consider the case where the detector also learns to distinguish between vortices and anti-vortices as two separate classes. Furthermore, since the sizes of vortices across the simulated images do not vary significantly, we focus on predicting the position of each vortex core rather than the full bounding boxes. If necessary the size of a vortex core can be determined by calculating the healing length of the condensate.

\begin{figure}[t]
    \centering
	\includegraphics[width=\textwidth]{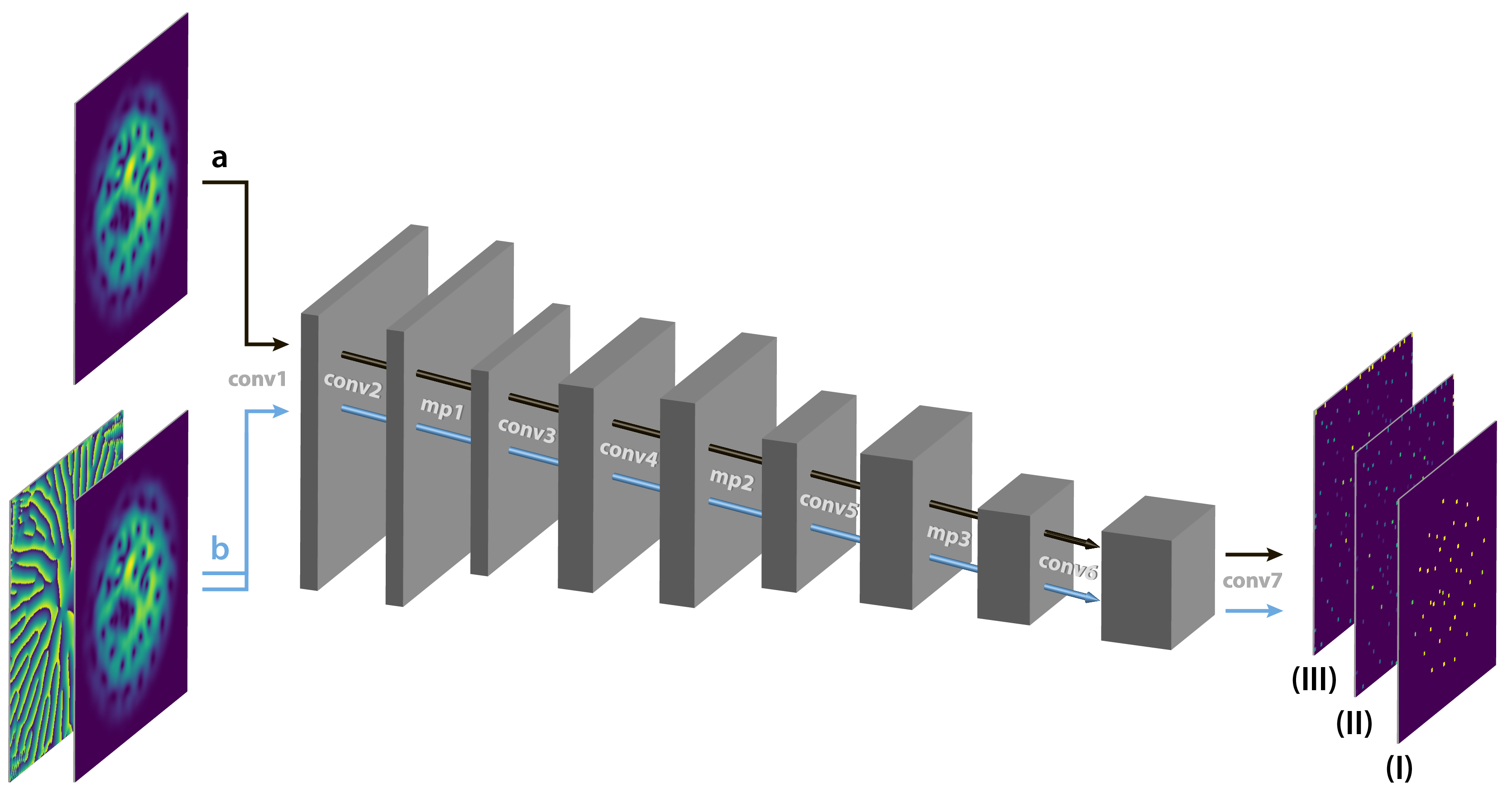}
	\caption{The network takes as input images of dimension $256\times 256$ with the condensate density alone (a), or the density and phase profile as two separate channels (b). The images are fed through 7 convolutional and 3 maxpool layers until the final layer outputs 3 matrices of dimension $64\times 64$. Each entry of the $64\times 64$ matrices is associated with a distinct $4\times 4$ cell in the original image and represents the probability of a vortex core being present inside the cell (I), and the scaled $x$ (II), and $y$ (III) position of the vortex within that cell.}
	\label{fig:02}
\end{figure}

The vortex detector takes as input gray-scale images $I \in [0,1]^{W \times H \times C}$ with equal width and height, $ W=H=256$, and a number of channels $ C=1,2 $ depending on whether the density profile or both density and phase profiles are provided to the neural network in two separate channels (see figure \ref{fig:02}). In principle, the output of the detector can assign a probability to each image pixel corresponding to whether the pixel represents a vortex core or not. However, due to the large dimensions of the input image, we divide it into a $\frac{W}{R} \times \frac{H}{R} $ grid with $R=4$ such that each $4\times 4$ grid cell is responsible for detecting at most one object. We estimated the size of vortices in our data set and thus, ensured that the grid is chosen fine enough such that at most one vortex is present in any cell. The output $Y_{ijk}$ of the neural network is therefore a tensor of dimensions $64 \times 64 \times 3$ where the 3 channels correspond to the probability of a vortex core being present, and the scaled $x$ and $y$ positions of the core within its grid cell.

In the following, we denote the neural network prediction by $Y$ and the ground-truth label by $\hat{Y}$. The latter are obtained by a brute-force detection method described in detail in \ref{ap:data}. Our training and test data is comprised of both ground state and out-of-equilibrium configurations which are obtained through numerical simulations of the GPE (see equation \eqref{eq:01}) with parameter values sampled uniformly from the range $g\in [50,600]$ for the interaction strength and $\Omega\in [0.65,0.95]$ for the rotation frequency. The obtained density and phase profiles are normalized such that their pixels lie between $[0,1]$ before being input to the convolutional neural network (CNN). The architecture is composed of 7 convolutional layers and 3 maxpool operations (see figure \ref{fig:02}). The full details of the architecture, the training, and the chosen hyperparameters are provided in \ref{ap:training}. We use the ADAM optimizer \cite{ADAM} and a loss function given by
\begin{align}\label{eq:02}
\begin{split}
	L =& \sum_{\text{batch}} \sum_{ij} \bigg[ -w_1 \hat{Y}_{ij1} \log\left(Y_{ij1}\right) - (1-\hat{Y}_{ij1}) \log\left(1-Y_{ij1}\right) \\
	&\qquad\qquad+ w_2 \hat{Y}_{ij1} \left(\left(\hat{Y}_{ij2}- Y_{ij2}\right)^2 + \left(\hat{Y}_{ij3} - Y_{ij3}\right)^2\right) \bigg] , 
\end{split}
\end{align}
where $w_1, w_2$ are hyperparameters. The first term in the loss function is the weighted cross entropy loss responsible for learning the correct assignment of vortex probabilities to each grid cell. We found that giving a higher weight to learning positive predictions stabilizes training since otherwise the network often learned to detect no vortices at all, likely due to the sparsity of vortices within an image. The last term is a mean-squared error (MSE) loss for the $ x $ and $ y $ positions of a vortex. Note, that only those entries of $Y$ with an existing vortex core contribute to this part of the loss function while all other entries are ignored and in general have arbitrary values. For evaluating and comparing the performance of the object detector we use widely adopted metrics in the field of object detection such as precision, recall, average precision (AP), and the F1 score that we compute on the test data set. For their definitions we refer to \ref{ap:metrics}.

\section{Vortex detection using density only}\label{sec:density}

\begin{figure}[tp]
	\centering
	\hspace*{-32mm}
	\begin{subfigure}[b]{0.3\textwidth}
		\centering
		\includegraphics[width=\textwidth]{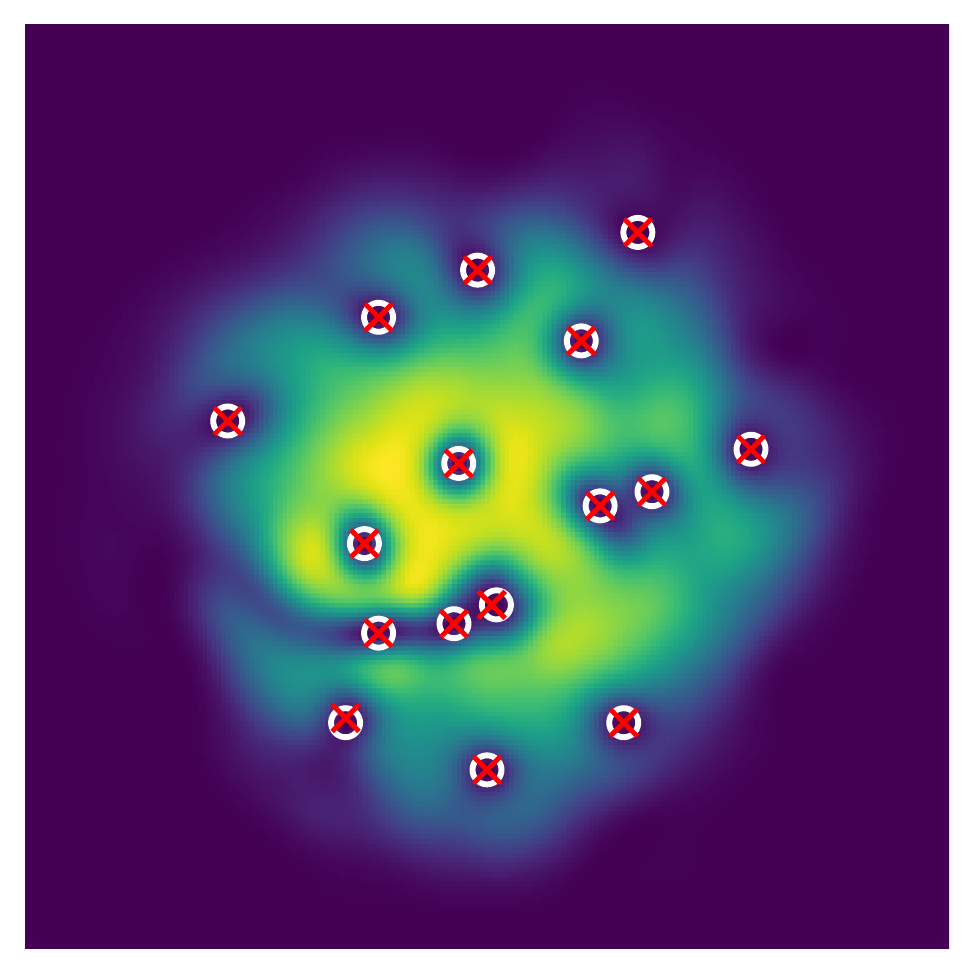}
		\caption{}
	\end{subfigure}
		\hspace*{1mm}
	\begin{subfigure}[b]{0.3\textwidth}
		\centering
		\includegraphics[width=\textwidth]{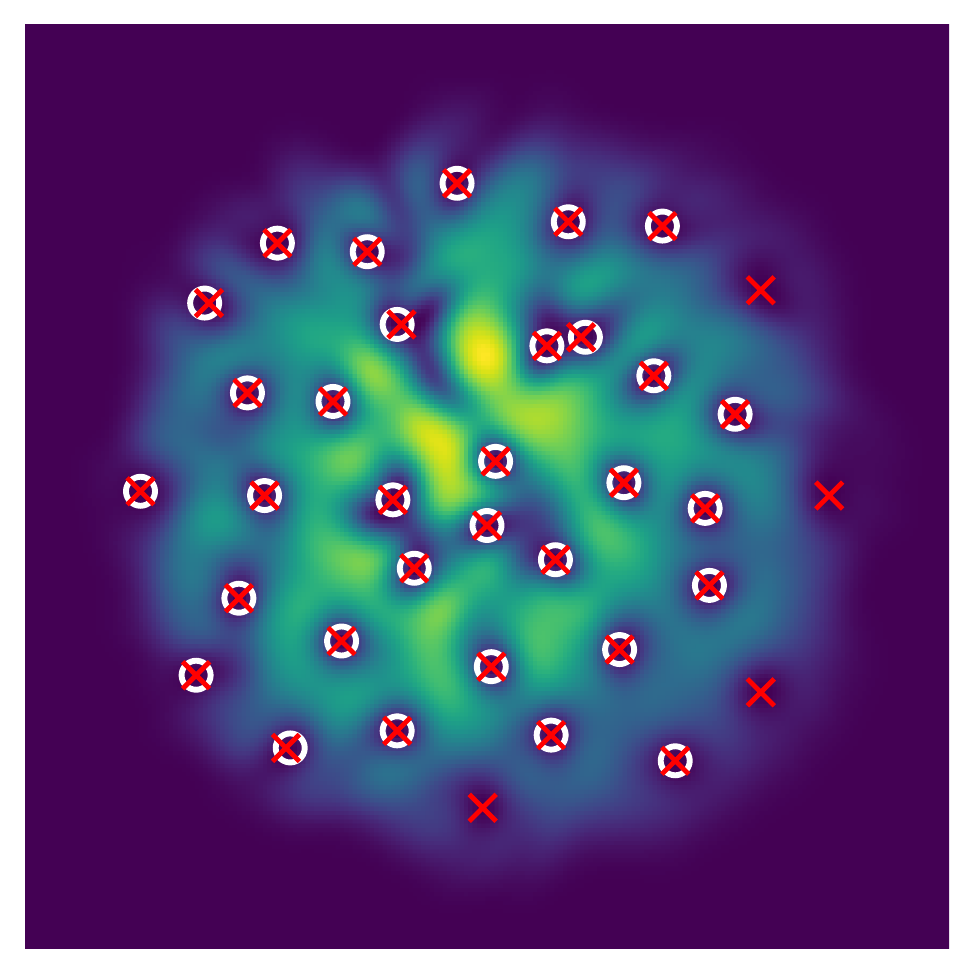}
		\caption{}
	\end{subfigure}
		\hspace*{1mm}
	\begin{subfigure}[b]{0.3\textwidth}
		\centering
		\includegraphics[width=\textwidth]{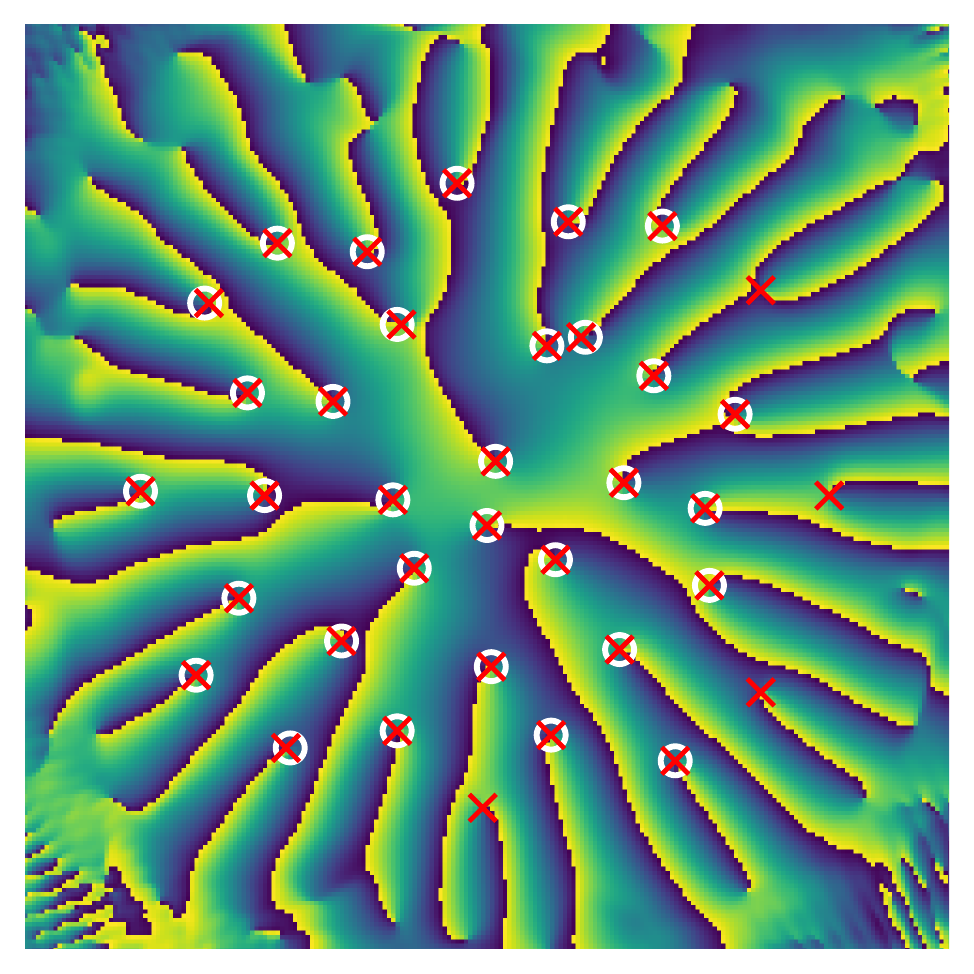}
		\caption{}
	\end{subfigure}
	\hspace*{-37mm}\hfill\\
	
	\hspace*{-32mm}
	\begin{subfigure}[b]{0.3\textwidth}
		\centering
		\includegraphics[width=\textwidth]{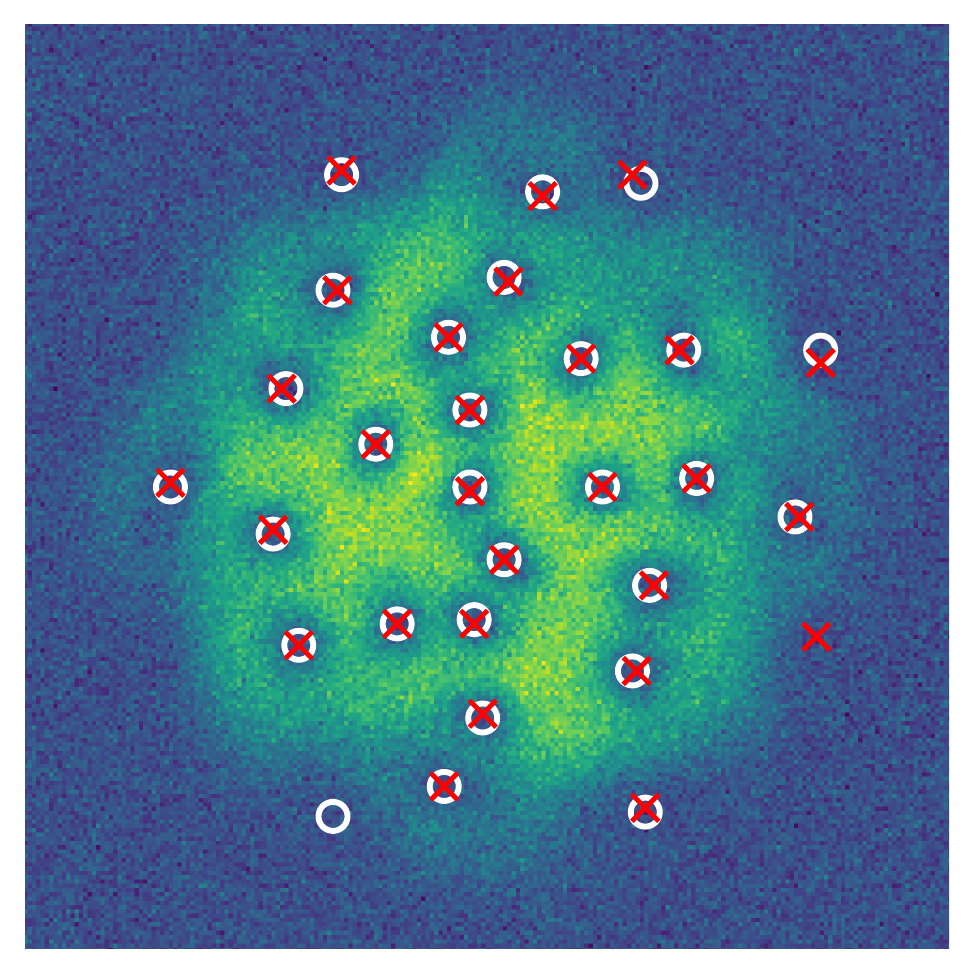}
		\caption{}
	\end{subfigure}
	\hspace*{1mm}
	\begin{subfigure}[b]{0.3\textwidth}
		\centering
		\includegraphics[width=\textwidth]{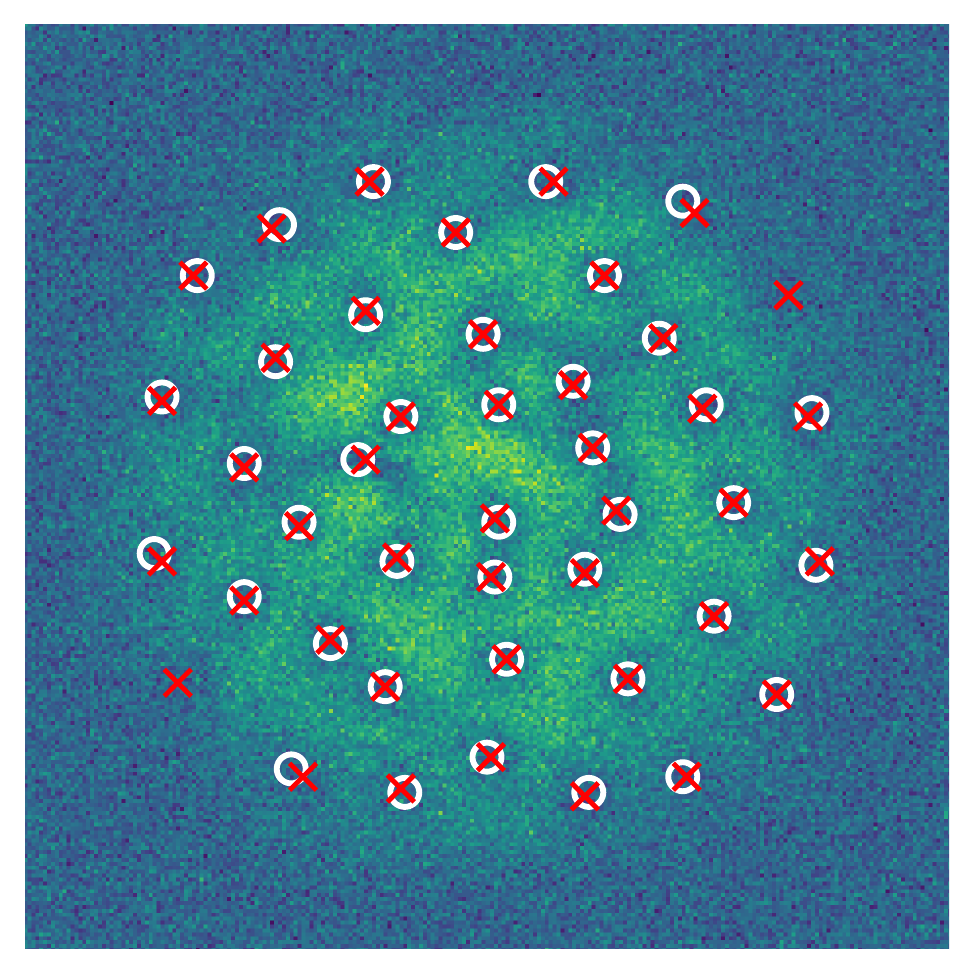}
		\caption{}
	\end{subfigure}
	\hspace*{1mm}
	\begin{subfigure}[b]{0.3\textwidth}
		\centering
		\includegraphics[width=\textwidth]{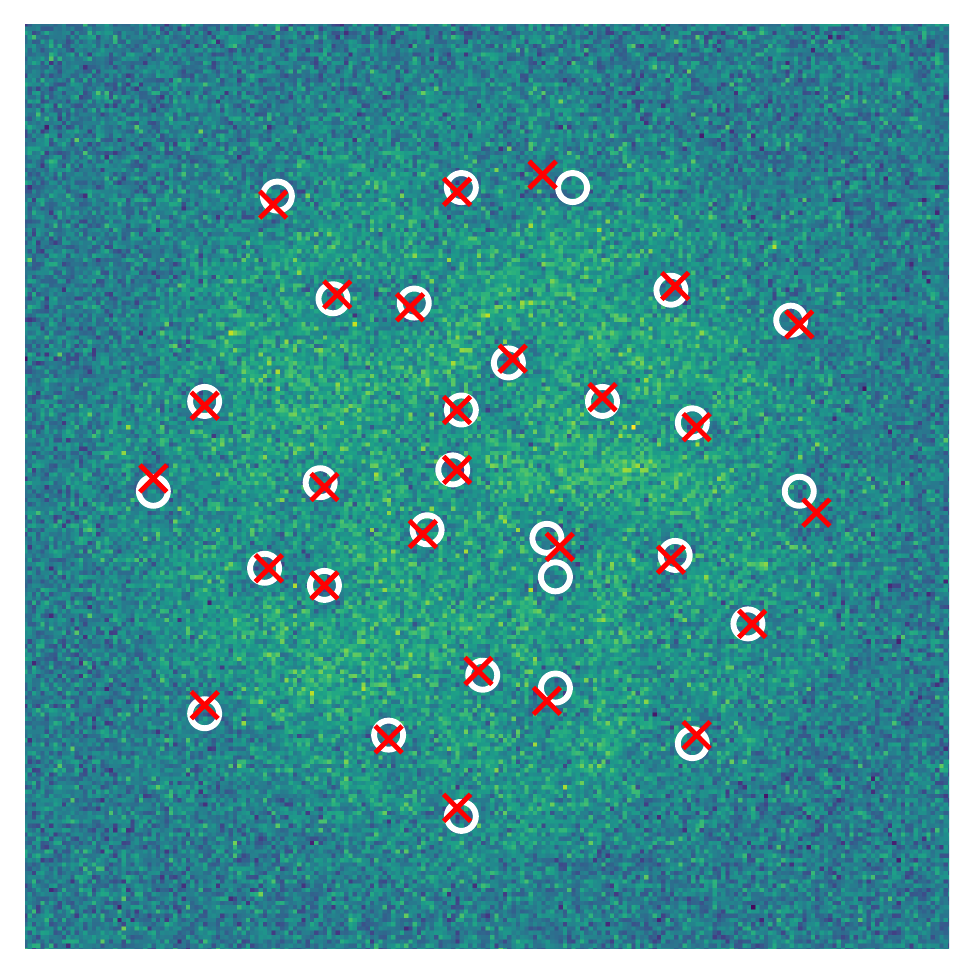}
		\caption{}
	\end{subfigure}
	\hspace*{-37mm}\hfill\\
	
	\hspace*{-32mm}
	\begin{subfigure}[b]{0.3\textwidth}
		\centering
		\includegraphics[width=\textwidth]{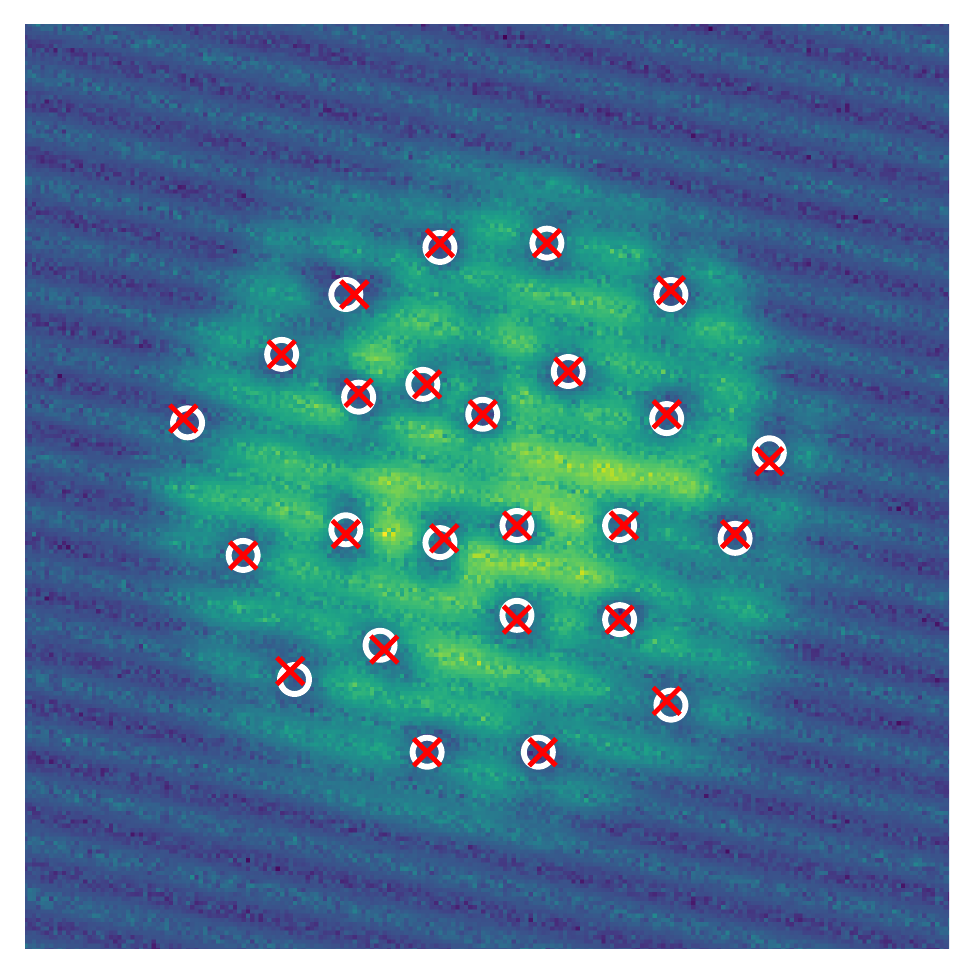}
		\caption{}
	\end{subfigure}
	\hspace*{1mm}
	\begin{subfigure}[b]{0.3\textwidth}
		\centering
		\includegraphics[width=\textwidth]{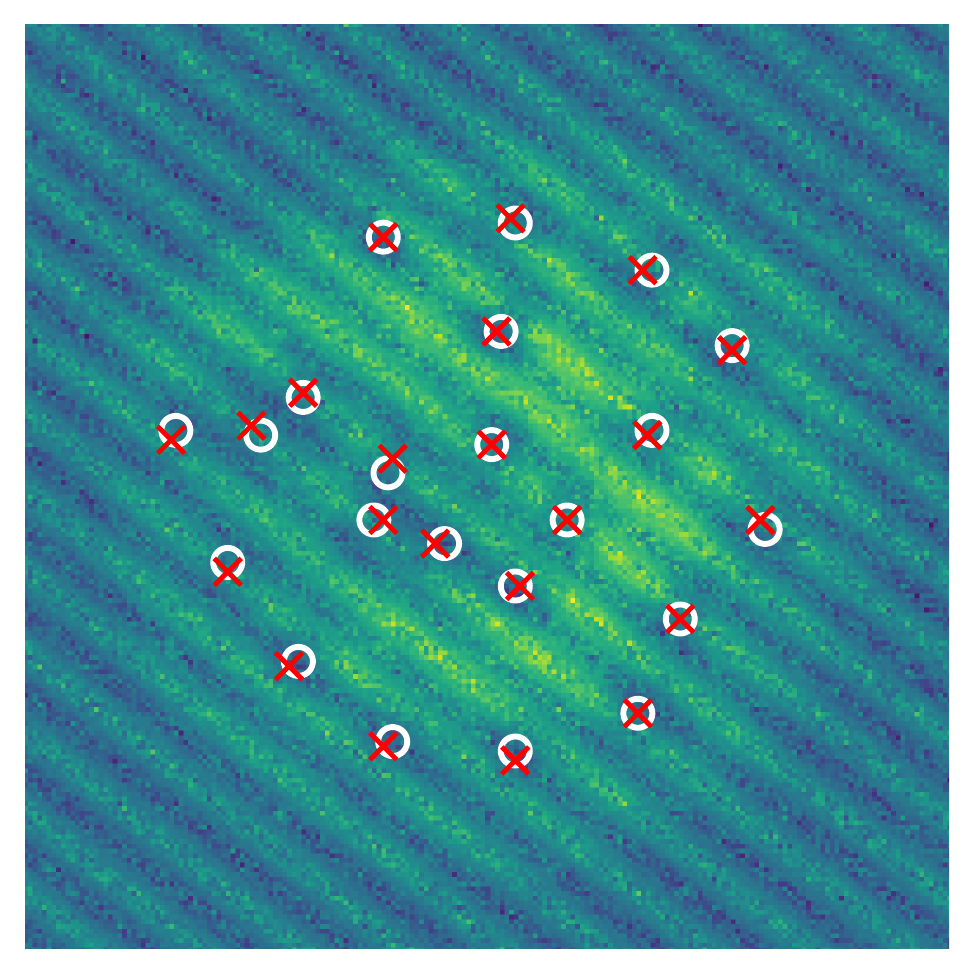}
		\caption{}
	\end{subfigure}
	\hspace*{1mm}
	\begin{subfigure}[b]{0.3\textwidth}
		\centering
		\includegraphics[width=\textwidth]{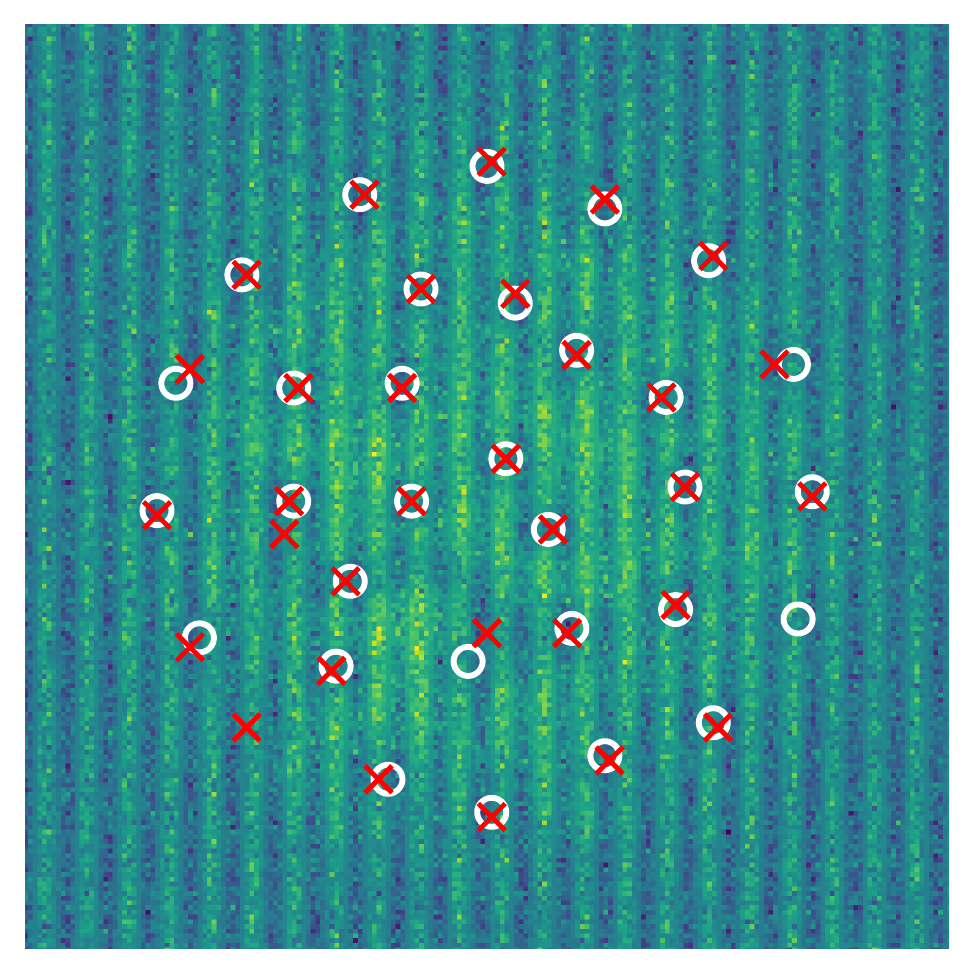}
		\caption{}
	\end{subfigure}
	\hspace*{-37mm}

	\caption{The locations of the vortex cores within each image are indicated by red crosses for the model prediction and by white circles for the ground truth obtained through the brute-force detection method. The CNN model was trained and tested only on BEC density images and therefore does not have access to the information encoded in the phase profile. (a) and (b) show two examples of BEC density configurations while (c) is the corresponding phase profile for the density image in (b) provided here as a guide for the eye. (d) - (f) display density distributions to which random Gaussian noise is added with growing standard deviations from left to right ($\sigma=0.1,0.2,0.5$). In (g) - (i) a sinusoidal modulation with Gaussian noise is added instead where the amplitude $A$ and the amount of noise increase from left to right ($A = 0.2, 0.5, 1.0$). Note that the pixels in the density images are normalized to lie between $[0,1]$ before including any noise and before being fed to the neural network.}
	\label{fig:03}
\end{figure}

First, we train the object detector directly on density images obtained from simulations with the GPE, i.e.~without any addition of noise. Figure \ref{fig:03}(a)-(b) show two representative density images with white circles corresponding to the ground truth and red crosses to the prediction of the trained model. Overall, we achieve a precision of $96.6\%$ and a recall value of $97.2\%$ on the test data (all other computed evaluation metrics can be inferred from table \ref{tab:01}). Precision and recall are calculated through comparison with the ground truth position obtained from the brute-force detection method which is not always accurate itself. Hence, our CNN likely performs better than the computed metrics.

While the network detects all vortices in figure \ref{fig:03}(a) with nearly perfect accuracy, we observe deviations from the ground truth label in the example shown in figure \ref{fig:03}(b). Here, the model detects additional vortices at the boundary of the condensate. However, the corresponding phase profile in figure \ref{fig:03}(c) features the characteristic phase winding at the location of the additional detections and hence these can be interpreted as vortices as well. In general we found that in most of the cases where the number of ground-truth detections and model detections differ, the missing/additional vortices lie at the boundary of the BEC and are often accompanied by a lower confidence probability. Note that the ground-truth labels were obtained using a brute-force detection algorithm which involves applying an arbitrarily chosen mask to the density images cutting off low density regions and therefore excluding any vortices that are not strictly within the BEC (see \ref{ap:data} for further details). Hence, during training the neural network also learns that vortices located in very low density regions should not be detected as such, however, it does not have access to the specific mask used in the brute-force detection algorithm. Therefore, it likely learns a slightly different density cutoff which gives rise to the additional detections in the test data. In figure~\ref{fig:09} of \ref{ap:layers} we visualize the output of one of the convolutional layers of the trained CNN which reveals information about the features learned by the model such as the specific masks applied by the CNN to differentiate regions within and outside the condensate.

\begin{table}[t!]
	\centering
	\begin{center}
		\begin{tabular}{ l | c | c | c | c }\hline
			 & Precision & Recall & AP & F1 \\ \hline \hline% \hline
			 Detection using density only (fig.~\ref{fig:03}) &  &  &  &  \\ 
			\qquad (a),(b) w/o noise & 96.6 & ~~97.2~~ & ~~95.1~~ & ~~96.9~~ \\ 
			\qquad \tab{(d)} \tab{weak Gaussian noise} & 93.9 & 93.8 & 91.1 & 93.9 \\ 
			\qquad \tab{(e)} \tab{moderate Gaussian noise} & 92.1 & 90.5 & 88.2 & 91.3 \\
			\qquad \tab{(f)} \tab{strong Gaussian noise} & 84.7 & 78.2 & 78.5 & 81.3 \\
			\qquad \tab{(g)} \tab{weak stripes} & 90.9 & 90.5 & 88.2 & 90.7 \\ 
			\qquad \tab{(h)} \tab{moderate stripes} & 88.4 & 88.3 & 83.4 & 88.4 \\ 
			\qquad \tab{(i)} \tab{strong stripes} & 85.0 & 83.9 & 78.8 & 84.5 \\\hline
			Detection using density and phase (fig.~\ref{fig:04}) &  &  &  &  \\ 
			\qquad \tabb{(a),(d)} \tabb{weak Gaussian noise} & 95.1 & 95.5 & 92.4 & 95.3 \\
			\qquad \tabb{(b),(e)} \tabb{moderate Gaussian noise} & 92.4 & 92.3 & 88.0 & 92.3 \\
			\qquad \tabb{(c),(f)} \tabb{strong Gaussian noise} & 78.0 & 74.7 & 69.4 & 77.2 \\ \hline
			Detection and classification (fig.~\ref{fig:05}) &  &  &  &  \\ 
			\qquad w/o noise, vortex/anti-vortex & 94.9 & 96.5 & 92.3 & 95.7 \\\hline
		\end{tabular}
	\end{center}
	\caption{Detector performance metrics (precision, recall, (mean) average precision (AP), and maximum F1 score) computed on the test data for each trained model (see \ref{ap:metrics} for their definitions). In the case of detection using BEC density images only, the Gaussian noise is added to the normalized density distributions with mean zero and standard deviations $\sigma=0.1$ (weak), $\sigma=0.2$ (moderate) $\sigma=0.5$ (strong). The stripe pattern was achieved by adding a sinusoidal modulation instead with amplitudes $A = 0.2$ (weak), $A = 0.5$ (moderate), $A = 1.0$ (strong). Finally, in the case of using both the density and the phase profiles as input to the CNN, the Gaussian noise was added directly to the real and imaginary parts of the wave function.}
	\label{tab:01}
\end{table}

To emulate experimental conditions we trained separate networks on images with two different sources of noise. The first type is Gaussian random noise with mean zero which is added to each pixel of the normalized condensate density images and mimics the measured density distributions in for example reference \cite{Folling05}. We trained three independent CNNs each with a different level of noise, i.e.~a different standard deviation ($\sigma = 0.1,0.2,0.5$), and plot the resulting predictions together with their ground-truth in figure \ref{fig:03}(d)-(f). As a another example of experimentally relevant noise we consider stripes in the density images which resemble the fringe patterns that can arise in absorption imaging due to unwanted interference effects \cite{Ness20,Song20}. To mimic this pattern we add a sinusoidal modulation to the density with randomly chosen direction and period. In addition, we add Gaussian random noise to the amplitude of the modulation itself. Figure \ref{fig:03}(g)-(i) show the corresponding density images together with the model prediction where the amplitude $A$ of the modulation and the amount of noise increases from left to right ($A = 0.2, 0.5, 1.0,\ \sigma = 0.2, 0.5, 1.0$). As expected, for both considered types of noise the performance of the detector deteriorates as the amount of noise increases which is also reflected by a smaller precision and recall value (see table \ref{tab:01}). In general we found that, as the noise grows, first only the predicted vortex positions become less accurate while for larger amounts of noise the model starts to entirely miss or mistakenly place vortices.

Note that although we have trained separate models for each different level of noise, we observed that each of the trained networks is able to generalize well to a different strength and type of noise which is crucial for real experimental situations where the amount of noise will likely change between measured images and experimental runs. For example, the model trained solely on strong Gaussian noise with $\sigma = 0.5$ achieved both a precision and recall of approximately $90\%$ on the test images containing a lower amount of noise and hence performs only slightly worse than the networks that have been directly trained on those data sets. The same model also performed well on images with weak stripes, however, for the case of moderate and strong stripes the model performance is considerably worse. This trend is however expected since the stripe pattern contains unique features that the model has not been exposed to during training. We also found that a network trained on a lower noise level generalizes to a certain extent to data involving more noise. For instance, the network trained on images with weak stripes achieves good accuracies on the test images with weak/moderate Gaussian and stripe noise with $F1$ scores over $80\%$, and only has difficulties locating vortices in images with a very large amount of noise. A summary of the computed evaluation metrics for the two examples discussed here is given in table \ref{tab:03} in \ref{ap:noise}.

Finally, in \ref{ap:ring} we show an example in which the model trained solely on images of BECs in a harmonic trap can also detect vortices in BECs in ring-shaped traps without any further training. We therefore expect the model to generalize to similar trapping potentials.

\section{Vortex detection using density and phase}\label{sec:phase}

\begin{figure}[t]
	\centering
	\hspace*{-32mm}
	\begin{subfigure}[b]{0.3\textwidth}
		\centering
		\includegraphics[width=\textwidth]{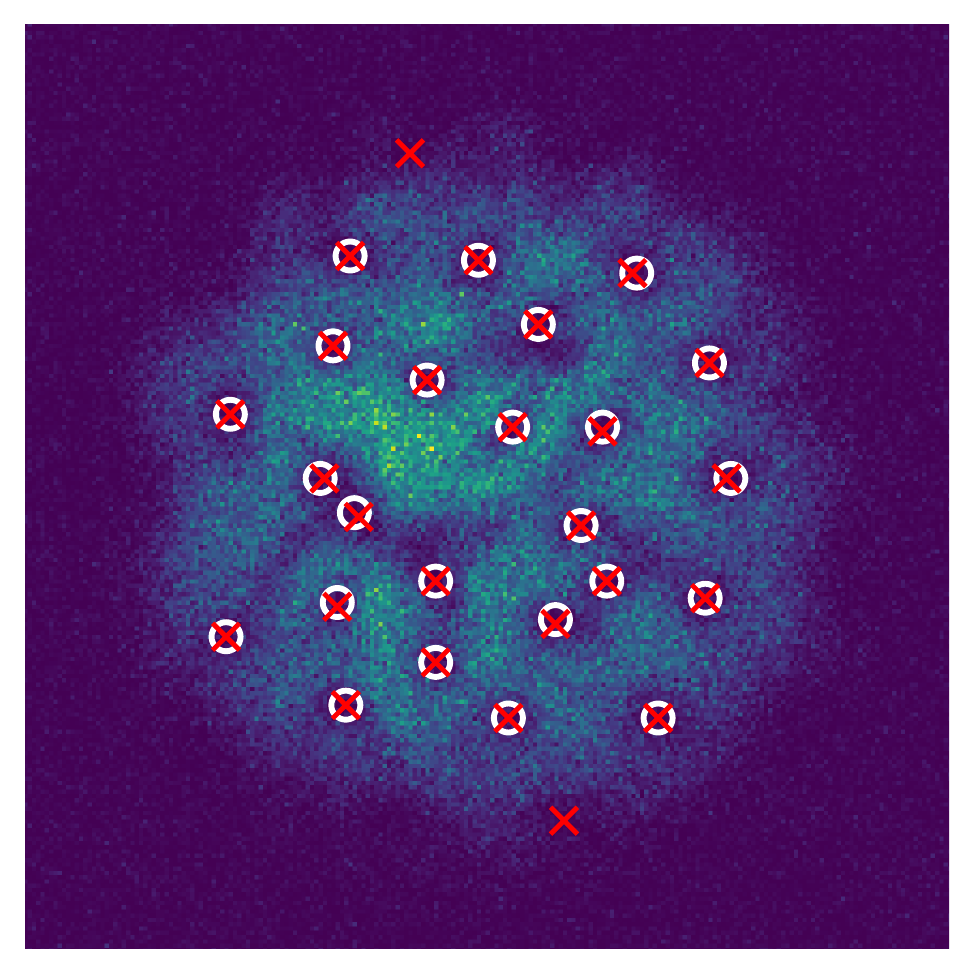} 
		\caption{}
	\end{subfigure}
		\hspace*{1mm}
	\begin{subfigure}[b]{0.3\textwidth}
		\centering
		\includegraphics[width=\textwidth]{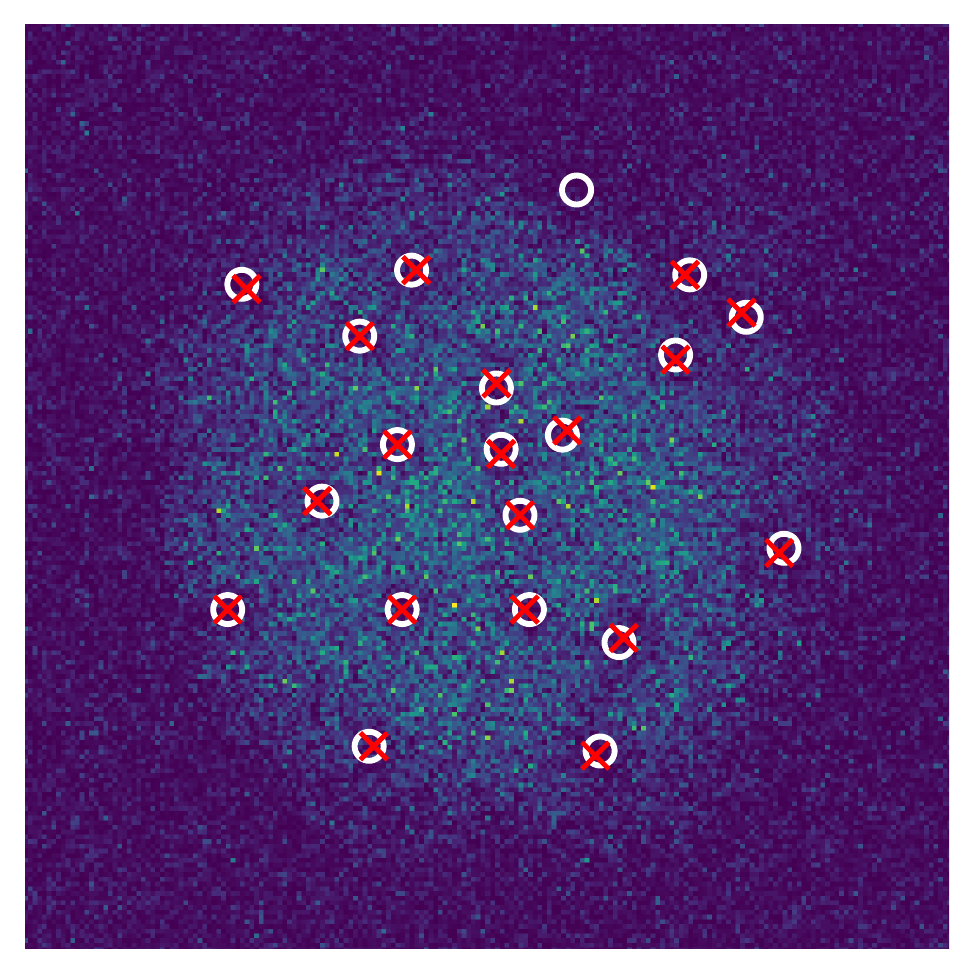} 
		\caption{}
	\end{subfigure}
		\hspace*{1mm}
	\begin{subfigure}[b]{0.3\textwidth}
		\centering
		\includegraphics[width=\textwidth]{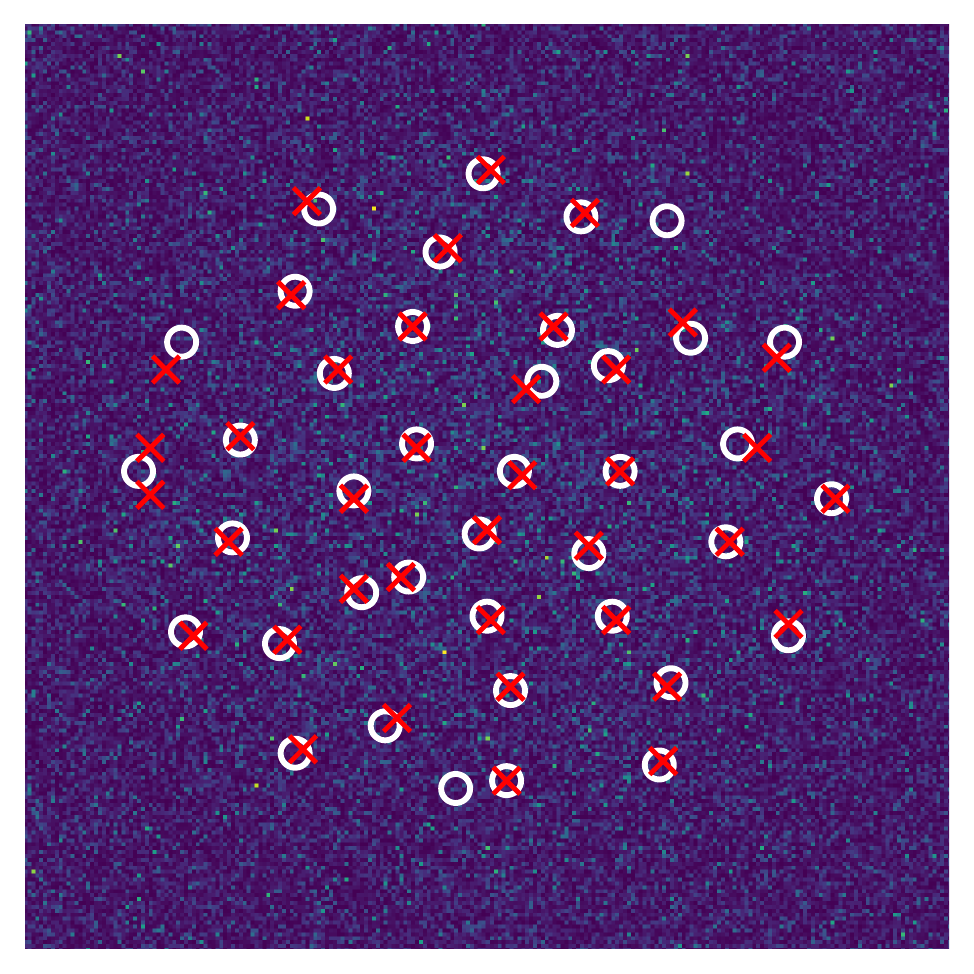} 
		\caption{}
	\end{subfigure}
	\hspace*{-37mm}\hfill\\
	
	\hspace*{-32mm}
	\begin{subfigure}[b]{0.3\textwidth}
		\centering
		\includegraphics[width=\textwidth]{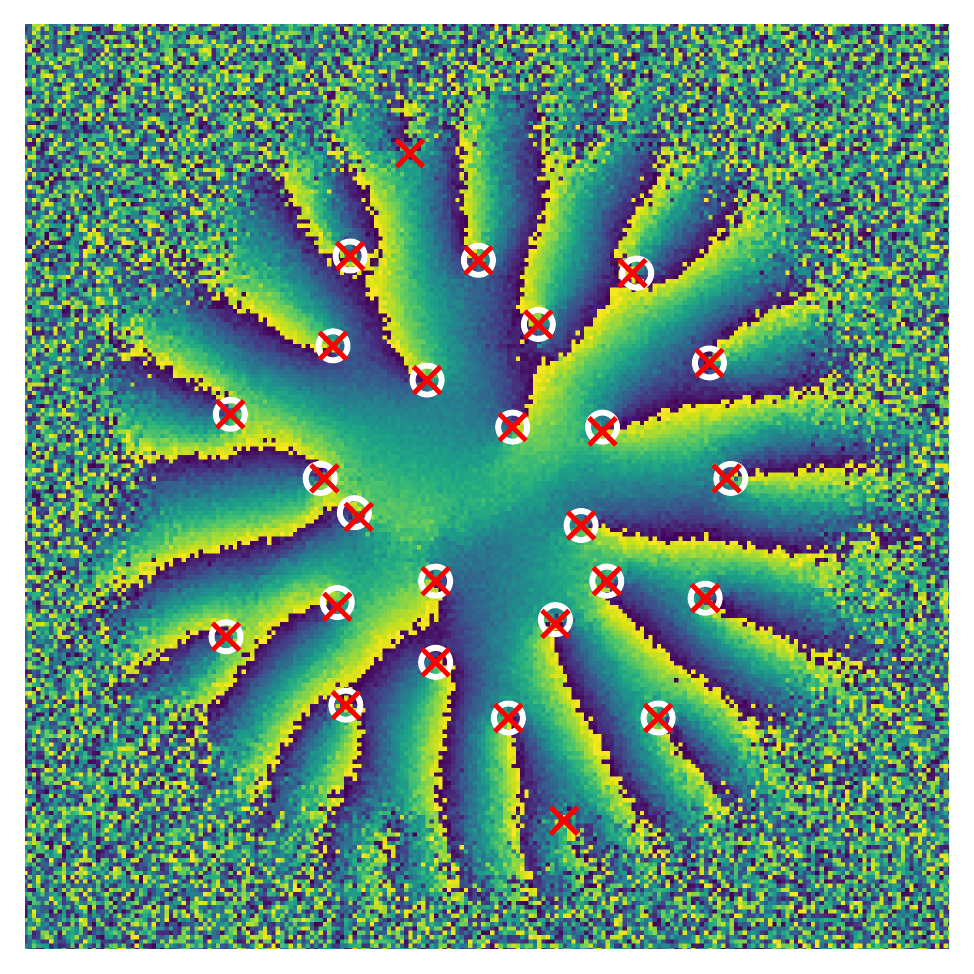}
		\caption{}
	\end{subfigure}
	\hspace*{1mm}
	\begin{subfigure}[b]{0.3\textwidth}
		\centering
		\includegraphics[width=\textwidth]{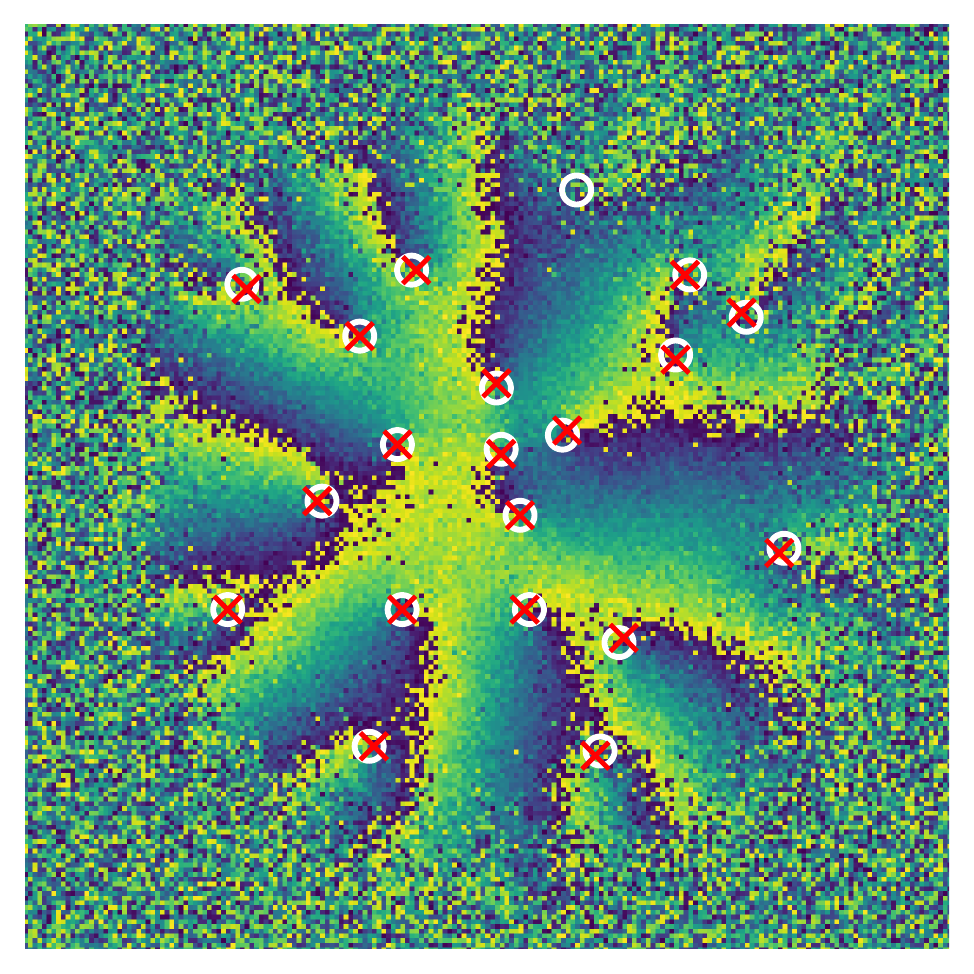}
		\caption{}
	\end{subfigure}
	\hspace*{1mm}
	\begin{subfigure}[b]{0.3\textwidth}
		\centering
		\includegraphics[width=\textwidth]{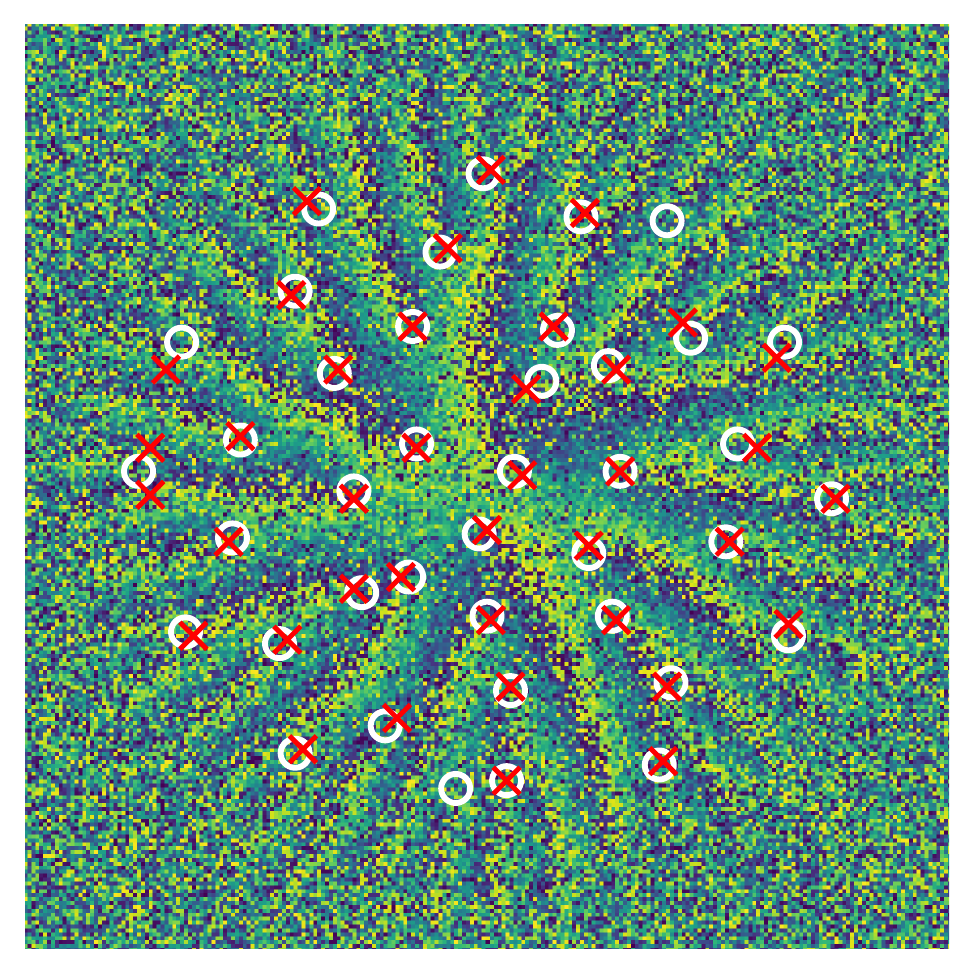}
		\caption{}
	\end{subfigure}
	\hspace*{-37mm}

	\caption{BEC density (a)-(c) and corresponding phase (d)-(f) configurations after adding Gaussian random noise to the real and imaginary parts of the condensate wave function. The amount of added noise increases from left to right. The model prediction is again denoted by red crosses while the ground truth is indicated by white circles. All images are normalized to lie between $[0,1]$.} 
	\label{fig:04}
\end{figure}

In numerical simulations of the GPE one has access to the full mean-field condensate wave function rather than just its density. Hence, we can provide both the density and the phase profile as input to the CNN in two separate channels in analogy to the three color RGB channels of a conventional image. In order to make the detection more challenging, we add Gaussian random noise to the real and imaginary part of the wave function which gives rise to the density and phase distributions depicted in figure \ref{fig:04}(a)-(c) and \ref{fig:04}(d)-(f) respectively. We train three models on different levels of noise and show the predicted vortex locations together with their ground truth in figure \ref{fig:04}. The achieved performance metrics can be read off table \ref{tab:01}. Especially for the case of weak noise, the detector performs very well with precision and recall values both above $95\%$ suggesting that the network exploits the additional information encoded in the BEC phase profile. In figure~\ref{fig:09} of \ref{ap:layers} we display the output of one of the CNN layers which also indicates that different features are learned depending on whether training is performed on just density or density and phase images.

\begin{figure}[t]
	\centering
	\hspace*{-32mm}
	\begin{subfigure}[b]{0.3\textwidth}
		\centering
		\includegraphics[width=\textwidth]{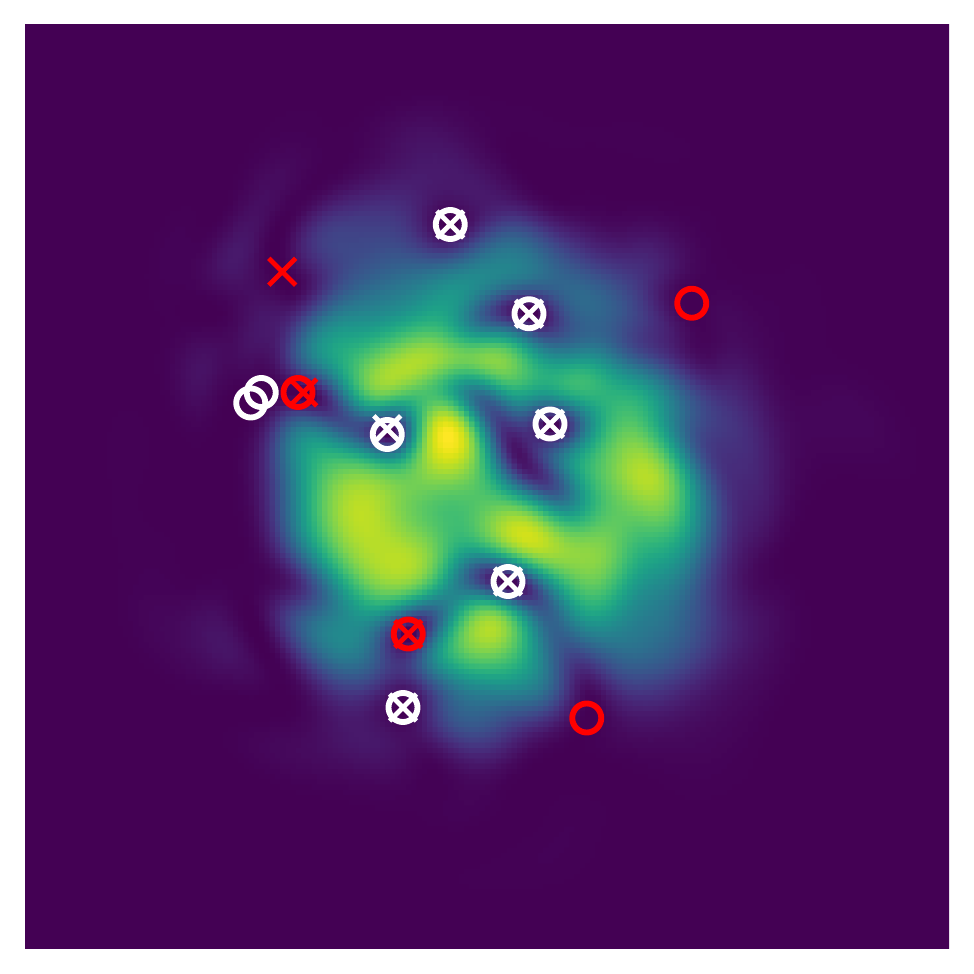} 
		\caption{}
	\end{subfigure}
		\hspace*{1mm}
	\begin{subfigure}[b]{0.3\textwidth}
		\centering
		\includegraphics[width=\textwidth]{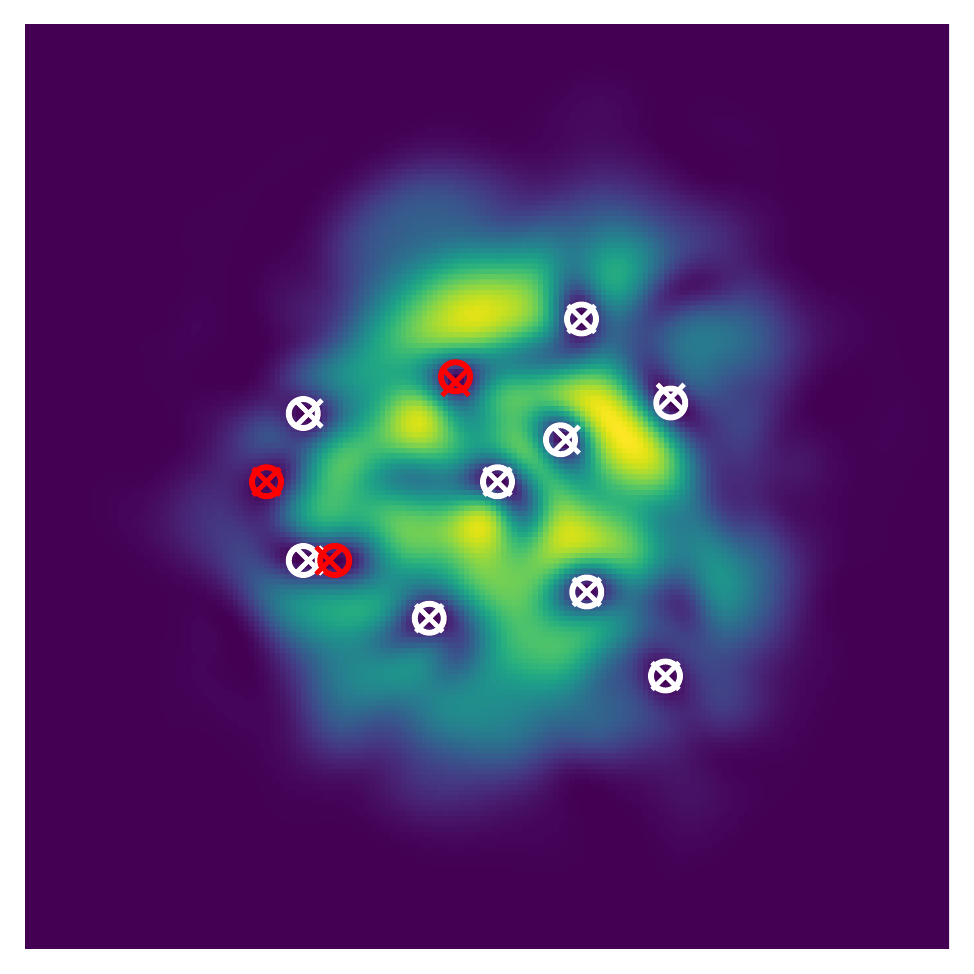} 
		\caption{}
	\end{subfigure}
		\hspace*{1mm}
	\begin{subfigure}[b]{0.3\textwidth}
		\centering
		\includegraphics[width=\textwidth]{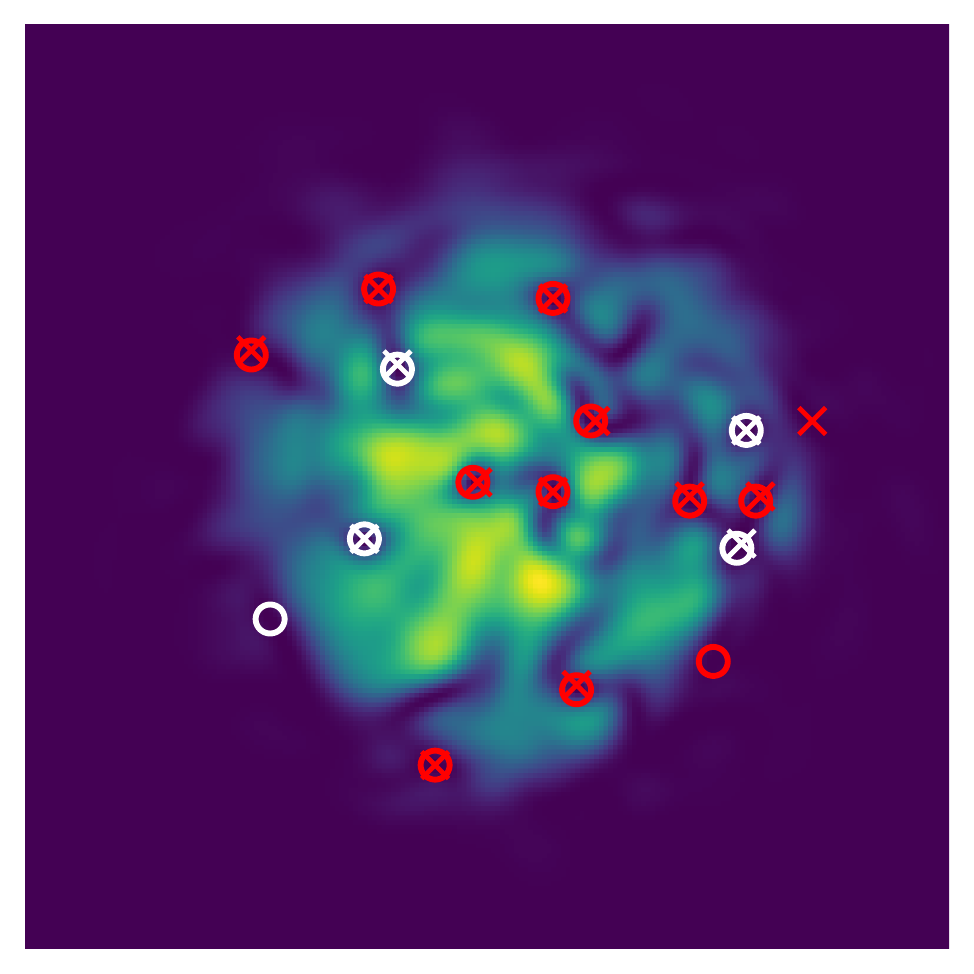} 
		\caption{}
	\end{subfigure}
	\hspace*{-37mm}\hfill\\
	
	\hspace*{-32mm}
	\begin{subfigure}[b]{0.3\textwidth}
		\centering
		\includegraphics[width=\textwidth]{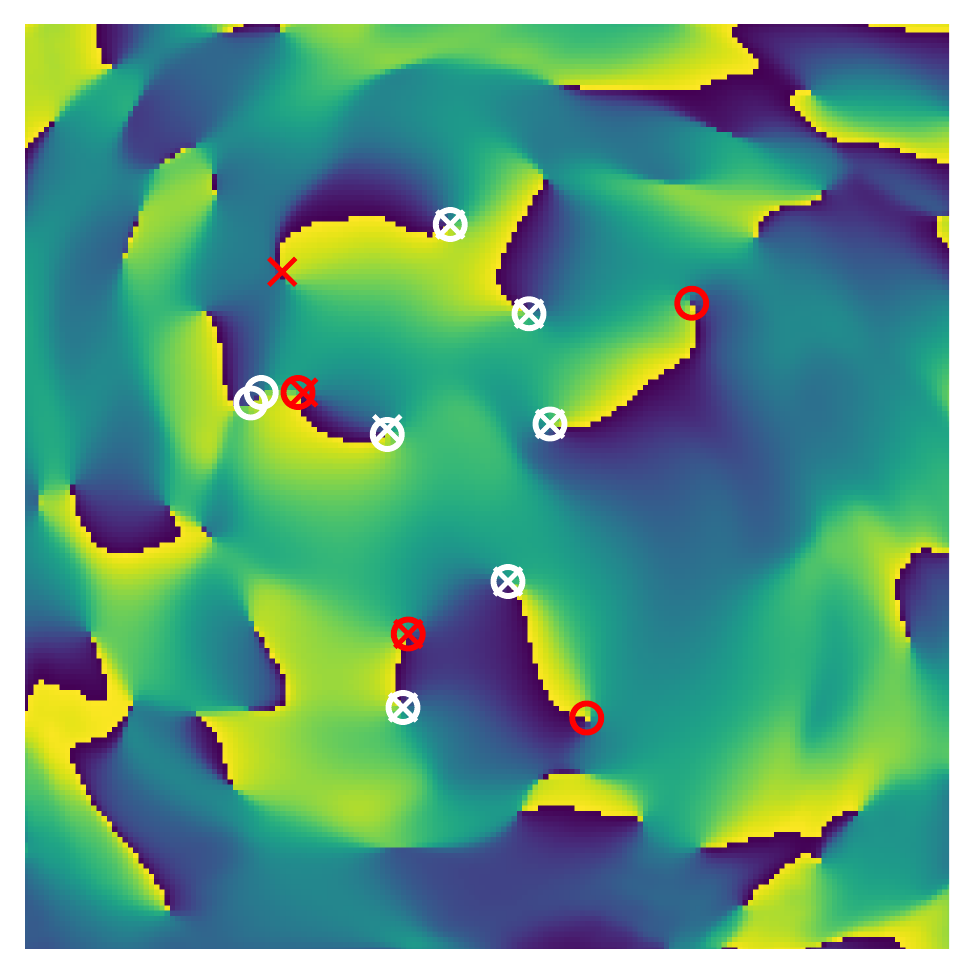}
		\caption{}
	\end{subfigure}
	\hspace*{1mm}
	\begin{subfigure}[b]{0.3\textwidth}
		\centering
		\includegraphics[width=\textwidth]{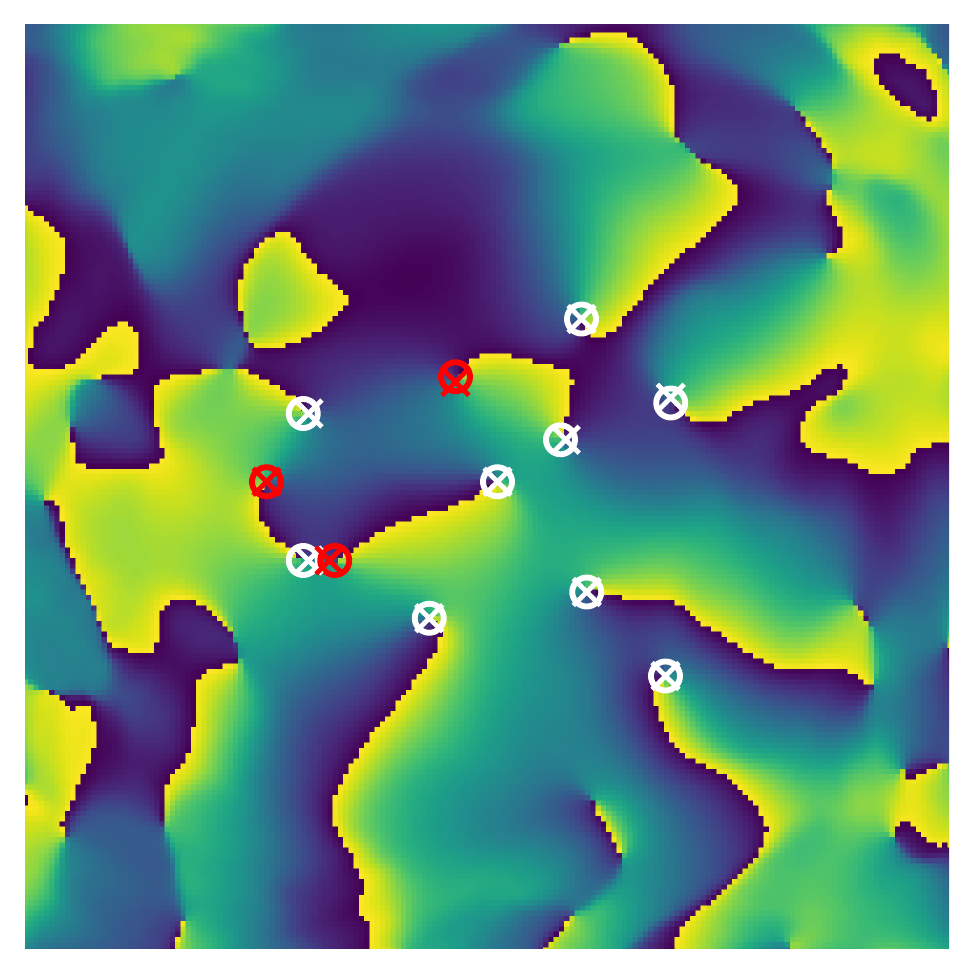}
		\caption{}
	\end{subfigure}
	\hspace*{1mm}
	\begin{subfigure}[b]{0.3\textwidth}
		\centering
		\includegraphics[width=\textwidth]{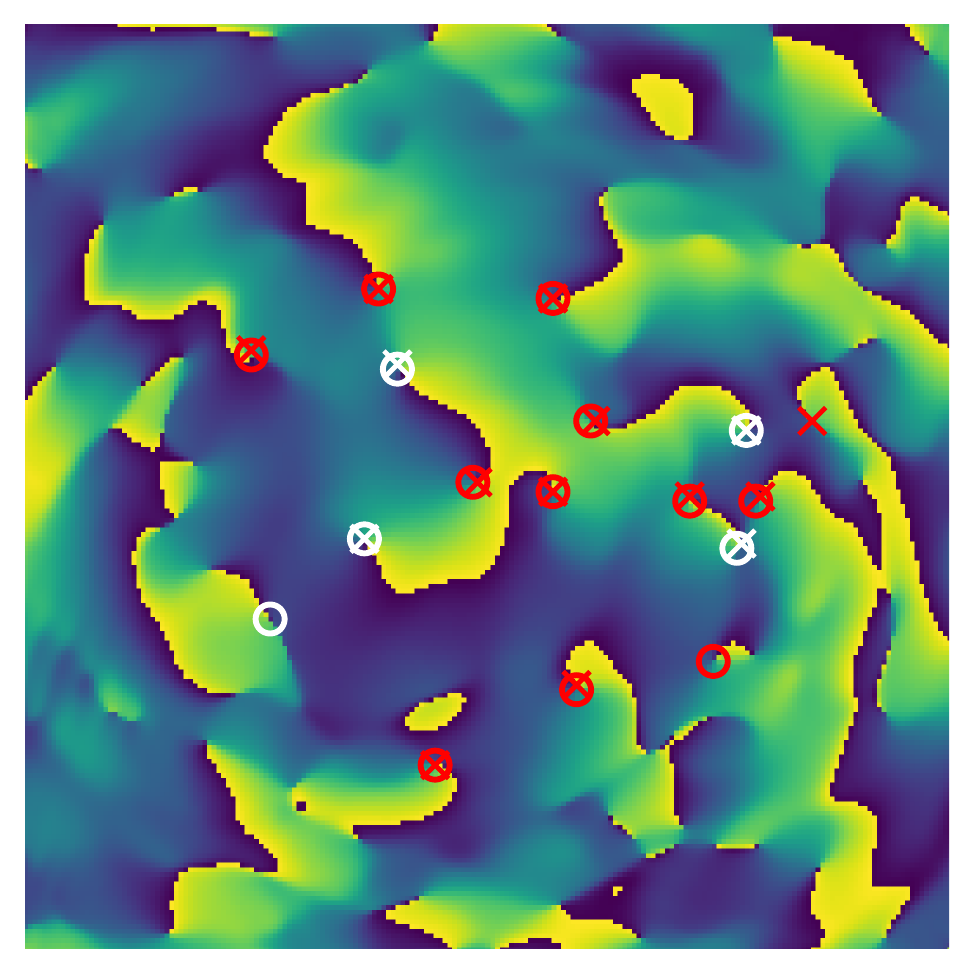}
		\caption{}
	\end{subfigure}
	\hspace*{-37mm}

	\caption{BEC density (a)-(c) and corresponding phase (d)-(f) configurations. The predicted vortex locations from the network are indicated by crosses and the ground truth by circles. The model was trained to also classify the circulation direction of a vortex with vortices depicted in red and anti-vortices in white.} 
	\label{fig:05}
\end{figure}

Furthermore, having access to the phase profile allows us to determine the direction of circulation through the sign of the phase winding. In the images shown in this paper (see for example figure \ref{fig:05}(d)-(f)) the circulation direction can be easily inferred by checking the direction of the color gradient when moving in a clock-wise loop around a vortex core, i.e.~whether the color changes continuously from yellow to blue or vice versa. Hence, we next train the network to also classify the sign of circulation for each vortex. The CNN is slightly altered to output 4 channels, the first two now corresponding to the probability of vortices and anti-vortices in a specific grid cell and the last two channels again contain the information about the precise location of a detected vortex. The loss function in equation \eqref{eq:02} is changed accordingly. Figure \ref{fig:05} shows three exemplary density and phase images where the model prediction is represented as crosses and the ground-truth as circles while vortices are depicted in red and anti-vortices in white. The model is able to accurately distinguish the circulation direction and in particular classifies all windings correctly for the images shown here. Moreover, it finds all vortices within the high-density region of the condensate which is also reflected by high precision and recall values as shown in table \ref{tab:01}.

\section{Discussion and Conclusion}

In this work we have presented a machine learning based vortex detector that can accurately predict the locations of vortices within two-dimensional BECs trapped in harmonic potentials. The machine learning model is based on a convolutional neural network (CNN) and takes as input either an image of the BEC density only or both, the BEC density and phase profiles. We first studied the experimentally more relevant case where only the condensate density is available and thus used for training and testing. Without any sources of noise the detector is able to reliably locate all vortices within an image. Moreover, the model performs well on non-equilibrium configurations that involve local density minima not corresponding to vortices. Hence, it learns to distinguish density minima arising due to vortices from those caused by other low energy excitations even in cases that are challenging for the human eye.

To simulate more realistic experimental conditions, we trained the network on density images with two different types of added noise that is, Gaussian noise and spurious stripe patterns arising due to unwanted optical interference effects \cite{Folling05,Ness20,Song20}. In either case, the achieved accuracy of the trained detector decreases with the amount of added noise as expected. However, overall the performance is still impressive given that it is nearly impossible to locate any vortices by eye in those images that contain a high level of noise. In contrast to the experimental setting, in numerical simulations both the density and phase profile of a BEC are available and therefore can be used to also distinguish the circulation direction of each vortex. In this case our detector learns to correctly classify the sign of circulation as well. Finally, the network is also able to accurately locate vortices in noisy configurations where Gaussian noise is added directly to the wave function itself. Due to the robustness of the detection against noise, it might be promising to train the detector on configurations generated by the stochastic GPE, which models BECs at finite temperatures \cite{Groszek20,Gautam14,Jackson09}.

We trained and tested the CNN using ground-truth labels obtained through a brute-force detection algorithm which already provides the position of vortices to a very good precision. This raises the question whether a machine learning approach is even necessary and advantageous. However, the brute-force detection algorithm has certain disadvantages: It involves searching for density minima and checking whether the conditions for a vortex, such as a $2\pi$ phase winding, are fulfilled which is neither efficient nor easily parallelizable. Furthermore, the algorithm does not achieve perfect accuracy itself, i.e.~it misses vortices or mistakenly places them at times. Additionally, the method heavily relies on the phase profile of the BEC for labeling out-of-equilibrium configurations, where local density minima may arise due to other excitations in the quantum system. However, in experiments the phase information is not easily accessible and therefore the algorithm cannot be straightforwardly applied in these settings. Finally, our brute-force method only works for simulated images without noise. While it is in general possible to improve the algorithm to also detect vortices in images with specific sources of noise, the implementation would be considerably more cumbersome. On the other hand, our machine learning based detector is robust to various sources of noise in the input data and does not rely on any hand engineered features.

The presented network can be trained in less than an hour on a single GPU and did not require elaborate hyperparameter tuning for any of the tasks considered here. Furthermore, the CNN is able to process several images in parallel and can therefore detect vortices in a large batch of input images fast given that the computation is performed on a GPU. For example, processing a batch of 100 images takes only on the order of milliseconds. The machine learning model can also be straightforwardly integrated with a GPU solver of the Gross–Pitaevskii equation which would eliminate the need to transfer data between CPU and GPU \cite{Schloss218}. As a possible next step the vortex detector could be combined with a tracking algorithm enabling the study of real-time dynamics of vortices in BECs such as in references \cite{Navarro13,James19,Koukouloyannis14}. It would also be interesting to compare the performance of our model to different architecture choices and object detection techniques, and we hope that our results can serve as a benchmark for further research on quantum vortex detection methods.

While the neural network has only been trained on images of BECs in a uniform harmonic trap, we found that the same model can detect vortices in ring-shaped traps without any additional training, and hence we expect that the detector also generalizes to other trapping geometries of similar symmetry. Moreover, we observed that a model trained on a particular strength and type of noise also worked well on different levels of noise. The promising generalization capabilities and the fact that the model performs well on density images alone with sources of noise and in the presence of spurious density minima suggests that the detector will be advantageous for experiments studying the dynamics of vortices in non-equilibrium states \cite{Reeves20,Johnstone19,Gauthier19}. Testing the trained model on real experimental data is therefore an interesting future direction of our work.

\section*{Acknowledgments}
This work was supported by OIST Graduate University and we are grateful for the help and support provided by the Scientific Computing
and Data Analysis Section of the Research Support Division at OIST. JP acknowledges funding from the JSPS KAKENHI Grant No.~20K14417.

\section*{Data availability}
The trained models from the main text as well as the code used for training and evaluating the network are available at \cite{github}. The data that support the findings of this study are available from the corresponding author upon reasonable request.

\appendix
\section{Training data generation}\label{ap:data}
Object detection belongs to the category of supervised learning tasks and therefore requires labeled training data. In order to include a variety of different vortex configurations for training, we use both ground states and non-equilibrium states generated within different parameter regimes. For each training example an interaction strength $g$ and a rotation frequency $\Omega$ are uniformly sampled in the range $g\in [50,600]$ and $\Omega\in [0.65,0.95]$. The ground state is obtained through imaginary time evolution using the split-step method \cite{Javanainen06}. To create out-of-equilibrium configurations containing both vortices and anti-vortices, we employ the method of phase imprinting \cite{Dobrek99} where between $4-7$ vortices are placed at random locations and with random circulation directions. We perform a short imaginary time propagation that simulates a small thermal relaxation of the system and a subsequent real-time evolution after which a snapshot of the condensate wave function is saved. The extracted condensate density and phase images comprise the training and test input data which include 1000 ground state samples and another 1000 out-of-equilibrium samples. The combined data set of 2000 images is randomly split into a training set containing 1600 images and a test set including the remaining 400 images.

The ground truth label for each image are the positions of all vortex cores inside the BEC. We obtain the $ x $ and $ y $ coordinates within pixel resolution through a combination of different techniques. We first apply a mask to the density and phase profiles cutting off regions outside the BEC. Hence, we ensure that only vortices strictly within the condensate are detected. As the cutoff threshold we choose $15\%$ of the maximum density $|\psi|^2$. Next, we find all local density minima within an image. For each local minimum we calculate the phase gradient along a closed loop centered at the minimum, check whether the slope equals $\pm 1$, and the loop adds up to $\pm 2\pi$, thus displaying the characteristic $2\pi$ phase winding. If all these conditions are met, the corresponding pixel position is stored in the list of labels. Note that this brute-force method of detecting vortices is not perfectly accurate and misses or mistakenly places vortices in a few cases. Hence, some of the labels used for training are corrupt, however, the overall excellent performance of the detector on the test data suggests that the training of the network is robust against the errors in the training set.

\begin{figure}[t]
	\centering
	\begin{subfigure}[b]{0.49\textwidth}
		\centering
		\includegraphics[width=\textwidth]{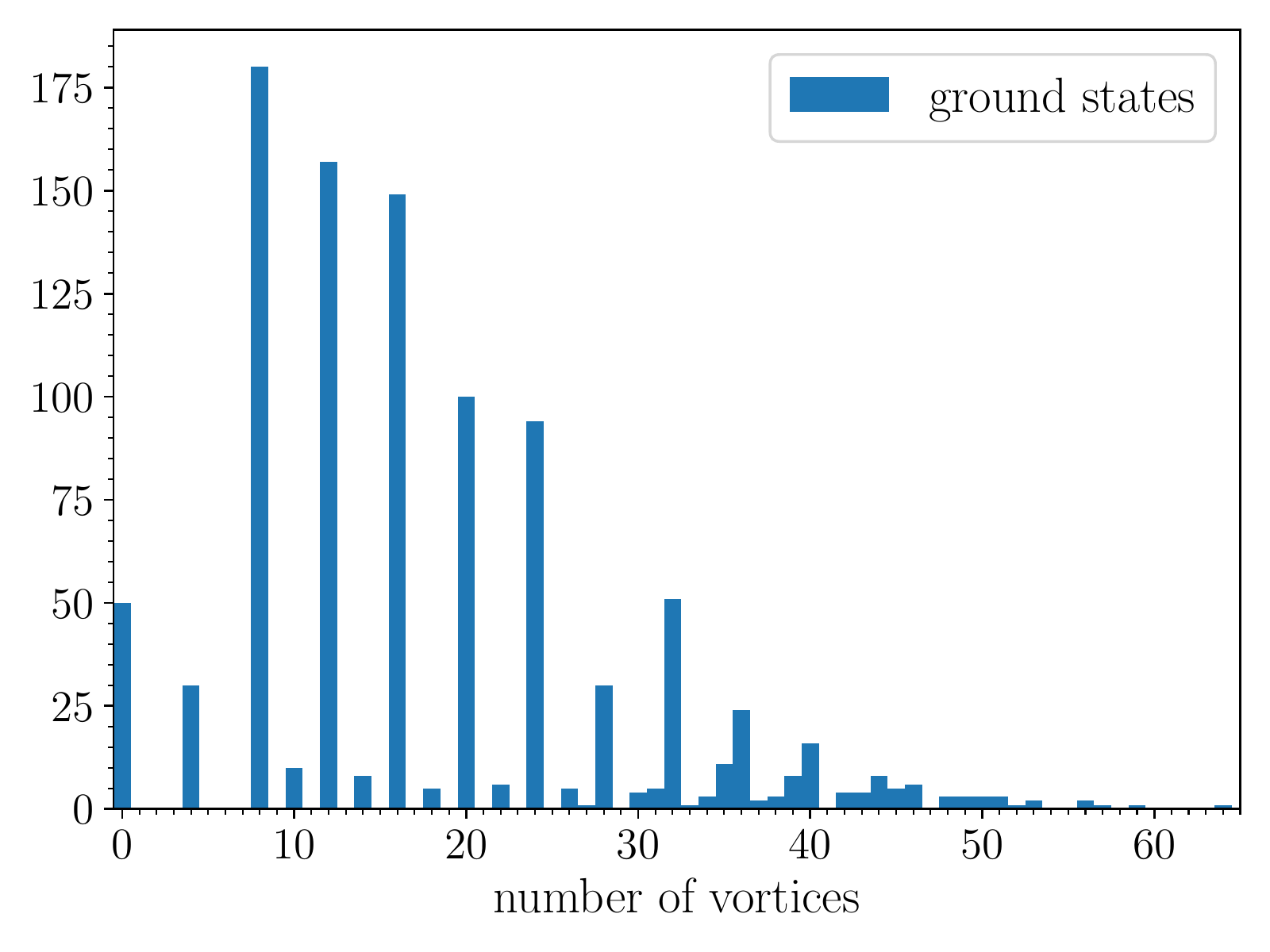} 
		\caption{}
	\end{subfigure}
		\hspace*{0.5mm}
	\begin{subfigure}[b]{0.49\textwidth}
		\centering
		\includegraphics[width=\textwidth]{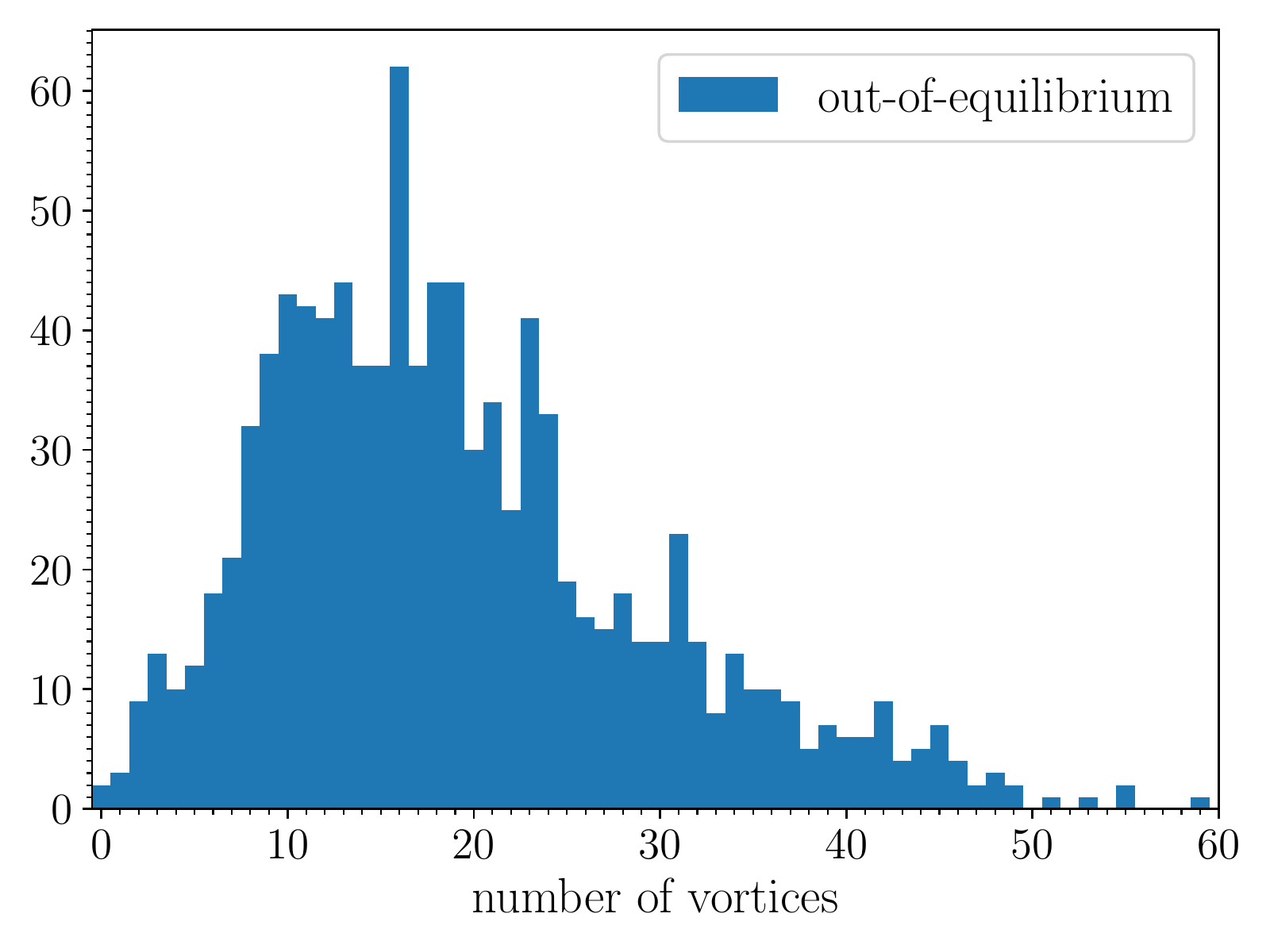} 
		\caption{}
	\end{subfigure}
	\caption{Distribution of the number of vortices across the training set for ground states (a) and out-of-equilibrium configurations (b). The combined distribution has a mean value of 18.7 vortices per image.} 
	\label{fig:06}
\end{figure}

We found that the number of vortices within a single image varied between 0 to 65 in our data set. The distribution of the number of vortices across all images is shown in figure \ref{fig:06}. The dependence of the number of vortices on the applied rotation frequency $\Omega$ is nonlinear, i.e.~for a large range of sampled rotation frequencies the number of vortices increases only slowly, while in the high frequency regime the number grows more rapidly \cite{Kasamatsu03}. The effect can be observed in the distributions which are asymmetric and slightly shifted towards a lower number of vortices with a mean value of 18.7 vortices per image. 

\section{Neural network architecture and training details}\label{ap:training}

The architecture of the neural-network based vortex detector is based on SlimNet, a convolutional neural network (CNN) specifically designed for detecting small objects \cite{Yang19}. It contains convolutional as well as maxpool layers as depicted in table \ref{tab:02}. The network takes as input images of size $256\times 256$ which can be either the density profile or the density and phase profiles provided in two separate channels. All convolutional layers are followed by relu activations except for the last layer, which uses a sigmoid activation instead. The output $Y$ of the network after the final convolutional layer is a $64\times64\times3$ tensor where each of the three channels corresponds to the probability of detection, and the scaled x, and y positions respectively. For example, an output $Y_{ij1} = 1$ would indicate that a vortex is present in the grid cell denoted by $ij$ and the precise position of the core within that grid cell can be read off by checking $Y_{ij2}$ for the $ x $ and $Y_{ij3}$ for the $ y $ coordinates. On the other hand, all grid cells where $Y_{ij1} = 0$ do not contain a vortex and therefore the second and third output channels can be ignored. In the case where the circulation direction of a given vortex is also classified, the output contains 4 channels with the first two representing the probability for a vortex and anti-vortex respectively. The final maxpool layer serves as a non-max suppression to eliminate multiple detections of the same vortex.

We implement the CNN and train it using Julia's machine learning library Flux \cite{Flux}. We use the ADAM optimizer \cite{ADAM} with a batch size of 100, a learning rate $\eta = 0.001$, and decay rates $\beta_1 = 0.9$ for the first and $\beta_1 = 0.999$ for the second momentum estimates. The weights in the loss function of equation \eqref{eq:02} are set to $w_1 = w_2 = 10$ and the network is trained for $100\sim 500$ epochs depending on the learning task. 100 epochs of training took on the order of $10$ minutes on a NVIDIA TITAN X Pascal GPU. A subsequent forward pass took 0.0025 sec for a single image and 0.009 sec for a batch of 100 images therefore allowing for real-time detection once the model is successfully trained.

\begin{table}[t!]
	\centering
	\begin{center}
		\begin{tabular}{ l | c | c | c | c }\hline
			Layer & Filter & Stride & Pad & Channels \\ \hline \hline% \hline
			conv 1 & $ 3 \times 3 $ & 1 & $ 1 \times 1 $ & $ 1/2^* \rightarrow 10 $ \\ 
			conv 2 & $ 3 \times 3 $ & 1 & $ 1 \times 1 $ & $ 10 \rightarrow 10 $ \\ 
			maxpool 1 & $ 2 \times 2 $ & 2 & & $ 10 \rightarrow 10 $ \\ \hline
			conv 3 & $ 3 \times 3 $ & 1 & $ 1 \times 1 $ & $ 10 \rightarrow 20 $ \\ 
			conv 4 & $ 3 \times 3 $ & 1 & $ 2 \times 2 $ & $ 20 \rightarrow 20 $ \\ 
			maxpool 2 & $ 2 \times 2 $ & 2 & & $ 20 \rightarrow 20 $ \\ \hline
			conv 5 & $ 3 \times 3 $ & 1 & $ 1 \times 1 $ & $ 20 \rightarrow 30 $ \\ 
			maxpool 3 & $ 2 \times 2 $ & 1 & & $ 30 \rightarrow 30 $ \\ \hline
			conv 6 & $ 3 \times 3 $ & 1 & $ 1 \times 1 $ & $ 30 \rightarrow 40 $ \\ 
			conv 7 & $ 1 \times 1 $ & 1 &  & $ 40 \rightarrow 3/4^* $ \\ \hline
			maxpool 4 & $ 3 \times 3 $ & 1 & $ 1 \times 1 $ & $ 3/4^* \rightarrow 3/4^*$ \\ \hline
		\end{tabular}
	\end{center}
	\caption{Neural network architecture. The number of input and output channels (marked with a star) differ between learning tasks. The final maxpool layer serves as non-max suppression.}
	\label{tab:02}
\end{table}

\section{Evaluation metrics}\label{ap:metrics}

To evaluate and compare the performance of the vortex detector for each different learning task we use standard metrics in the field of object detection such as precision and recall \cite{Xiao20}. Precision describes how many of the detections within an image are accurate, while recall quantifies how many of the actual objects in an image are detected. Denoting true positives by $ TP $, false positives by $ FP $, and false negatives by $ FN $, precision and recall are defined through
\begin{align}
		\text { Precision }&=\frac{TP}{TP+FP}\ ,\\
		\text { Recall }&=\frac{TP}{TP+FN}\ .
\end{align}
The machine learning model outputs the probability for a vortex to be present in each grid cell. A confidence threshold is used to discard low probability detections and label high confidence predictions as positives $P$. Conventionally, a larger confidence threshold increases precision while decreasing recall and vice versa. For each different detection task we calculate an optimal confidence threshold that maximizes the harmonic mean of precision and recall, which we describe further below. To distinguish true positives from false positives, the object detection community usually computes the intersection over union (IoU) given by the ratio of the intersection area and union area of the detected bounding box and the ground-truth bounding box. If the IoU is larger than a predefined value, the detection is considered to be a true positive $TP$ while it is labeled as a false positive $FP$ otherwise. In our case the vortices across images are of similar sizes and therefore we can use a simple distance measure between the detected vortex position and the ground-truth position instead of the IoU. We choose the pixel-wise euclidean distance and an arbitrary distance thresholds of $\sqrt{5}$. Hence, all detections that are within $ \sim 2 $ pixels of their ground-truth position are identified as positives.

\begin{figure}[t]
	\centering
	\begin{subfigure}[b]{0.49\textwidth}
		\centering
		\includegraphics[width=\textwidth]{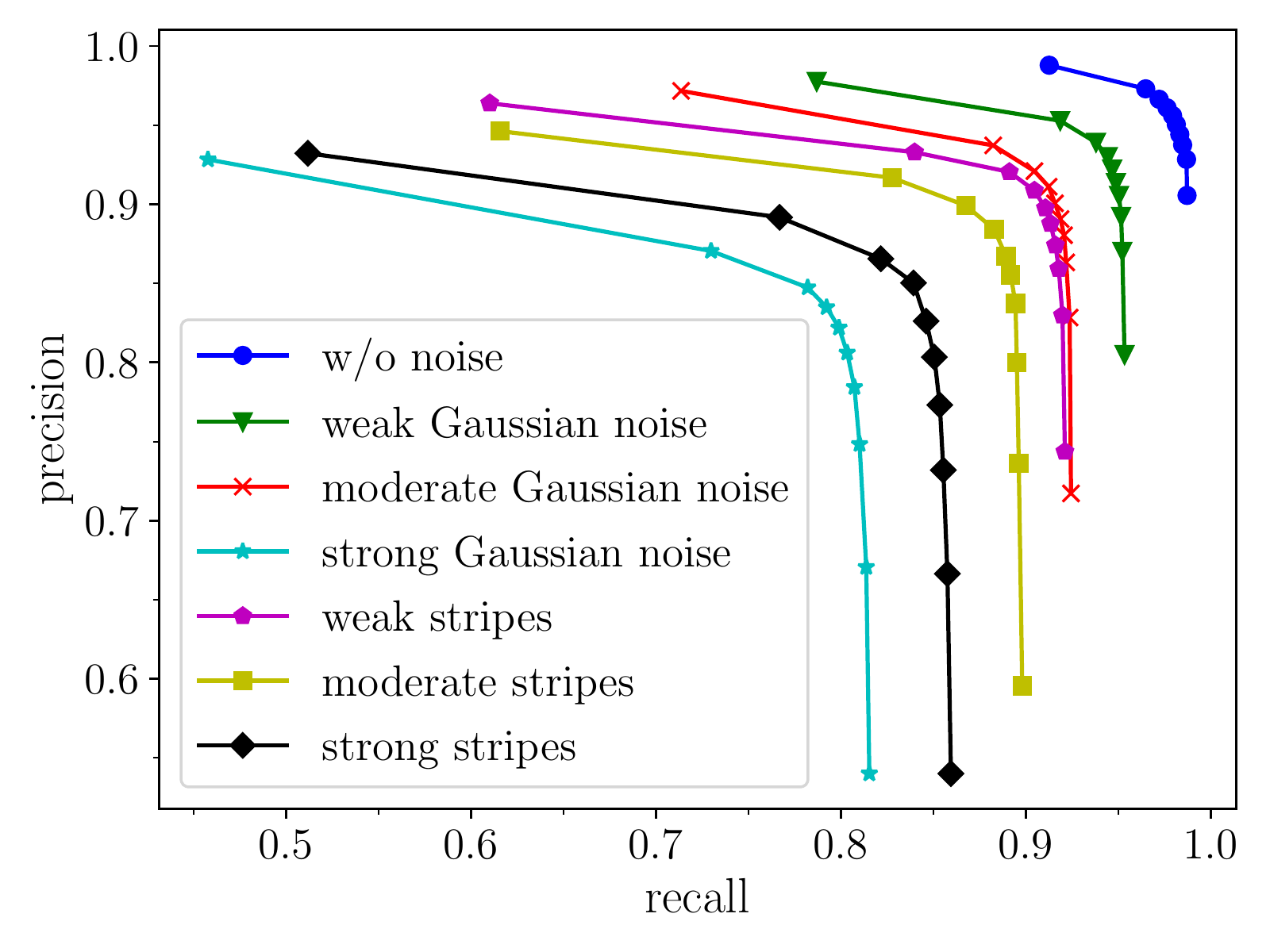} 
		\caption{}
	\end{subfigure}
		\hspace*{0.5mm}
	\begin{subfigure}[b]{0.49\textwidth}
		\centering
		\includegraphics[width=\textwidth]{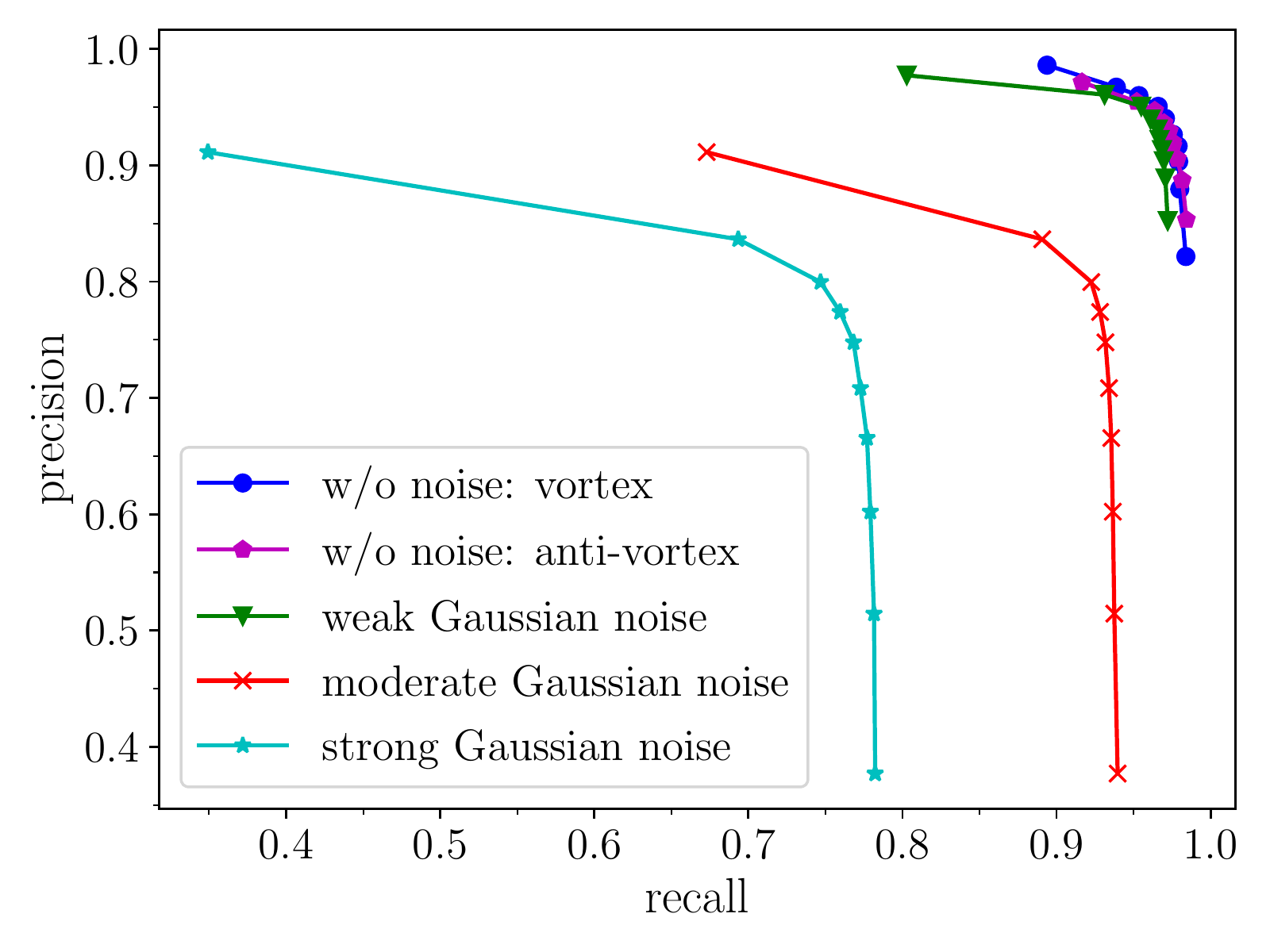} 
		\caption{}
	\end{subfigure}
	\caption{Precision - recall curves for the training tasks of Section \ref{sec:density} (a) and Section \ref{sec:phase} (b). Precision/recall values are calculated for 10 different confidence thresholds in the equally spaced interval between [0.05, 0.95]. The precision/recall values corresponding to the highest F1 score are shown in table \ref{tab:01} of the main text together with the maximum F1 score and the average of the precision values (AP).}
	\label{fig:07}
\end{figure}

As mentioned before there exists a trade-off between precision and recall controlled by the value of the confidence score threshold. To examine the performance of the model across different confidence thresholds, we plot precision against recall for 10 different threshold values in figure~\ref{fig:07}. Each curve of a different color corresponds to one of the separately trained models considered in the main text. From each point on a curve we can determine the F1 score, i.e.~the harmonic mean between precision and recall
\begin{equation}
	\text{F1}=2 \times \frac{\text {Precision} \times \text {Recall}}{\text {Precision}+\text {Recall}}\ .
\end{equation}
The optimal confidence threshold corresponds to the maximum F1 score which is used for generating all labeled plots within this paper. Furthermore, we calculate the average precision (AP) as the mean over the precision values $p(r)$. Finally, we also provide the precision and recall values computed at the optimal confidence threshold as another meaningful performance metric of the vortex detector. In the case of multi-class detection considered in Section \ref{sec:phase} the AP and F1 scores are first calculated separately for each class and then averaged to obtain mean average precision and a mean F1 score.

\section{Visualization of the CNN layers}\label{ap:layers}

\begin{figure}[tp!]
	\centering
	\begin{subfigure}[b]{0.75\textwidth}
		\centering
		\includegraphics[width=\textwidth]{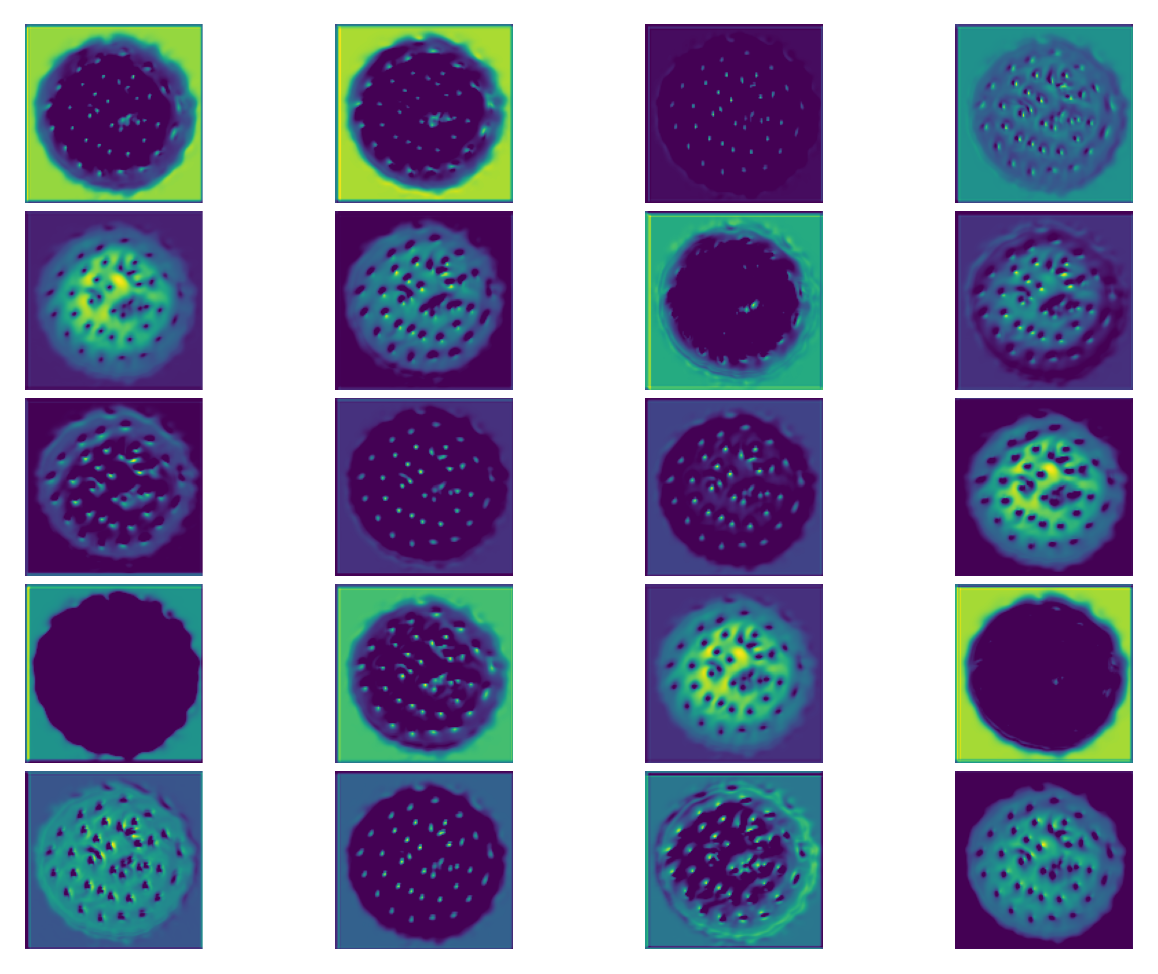}
		\caption{}
	\end{subfigure}\hfill\\
	\begin{subfigure}[b]{0.75\textwidth}
		\centering
		\includegraphics[width=\textwidth]{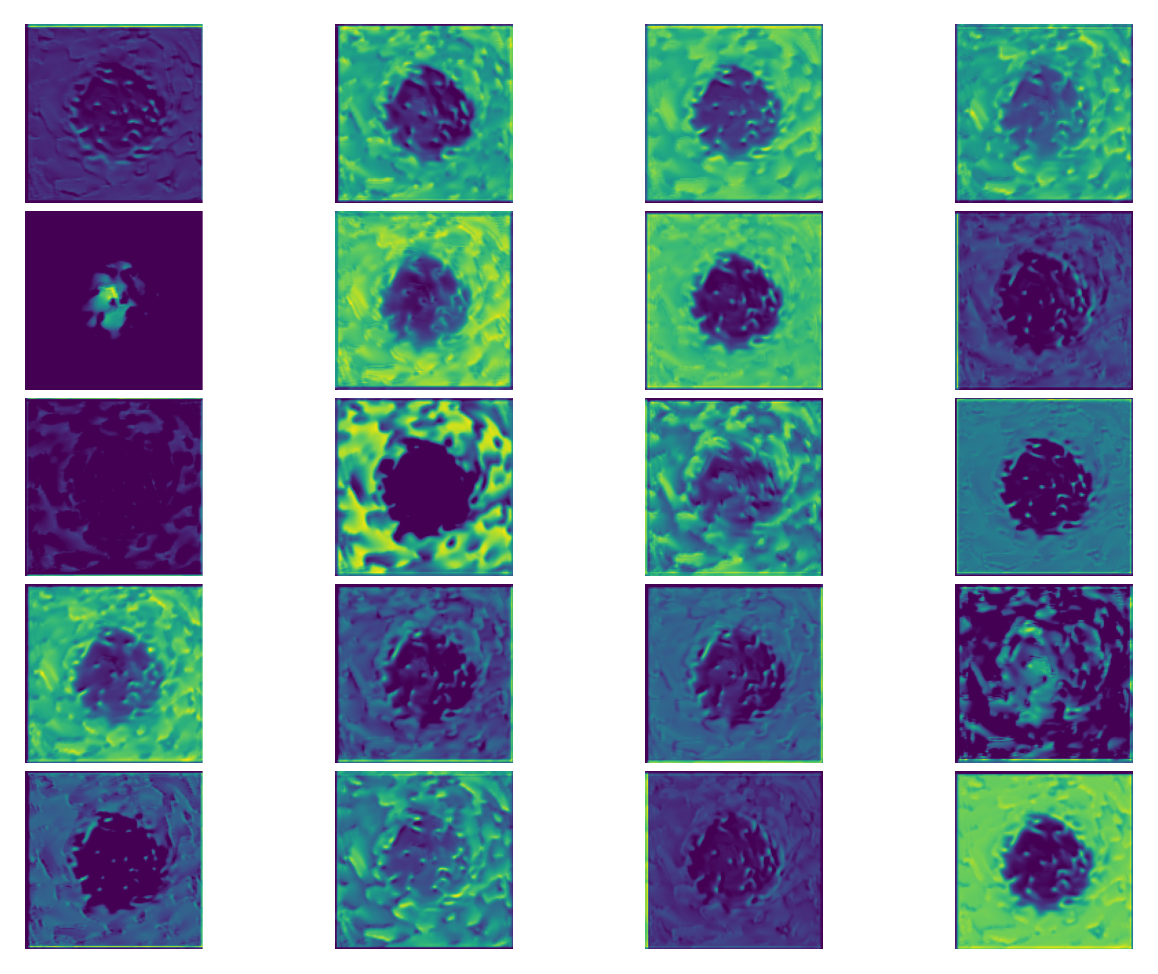}
		\caption{}
	\end{subfigure}
	\caption{Example output after the 4th convolutional layer of the CNN when training only on density images (a) or on density and phase images (b).} 
	\label{fig:09}
\end{figure}

To elucidate the inner workings of the trained CNN we visualize the output after the fourth convolutional layer for a particular input image in figure \ref{fig:09}. Each image corresponds to the output of one of the 20 channels and gives information about the features learned by the network. Figure \ref{fig:09}(a) shows the feature maps after training on only BEC density images. Here, the network clearly learns to separate the condensate from the background by applying different masks. It also detects all density minima within the condensate, however, each channel seems to focus on slightly different characteristics such as the depth, size, or shape of a minima. On the other hand, the feature maps plotted in figure~\ref{fig:09}(b) were obtained when training with phase and density images. In this case, an interpretation is less evident, but we can observe that the network takes advantage of the additional information supplied by the phase profile. Interestingly, we found that the output of CNN layers trained on noisy images did not differ significantly from the ones shown here. Therefore, the model also learns to de-noise the input if necessary and thus, ignores any spurious features contained in the noise itself.

\section{Generalization to different sources and levels of noise}\label{ap:noise}

The models considered in the main text of this paper were all trained on data sets containing a specific source and level of noise. However, real experimental images will not be subject to just a single source of noise and the amount of noise will likely vary between images. Therefore, we also test how well a model trained on a specific noise configuration generalizes to other types and levels of noise. As two examples we consider a model trained only on data with strong Gaussian noise or only on images with weak stripes. We show the computed performance metrics for both models tested on all different data sets in table \ref{tab:03}. The values suggest that the networks indeed generalize to unseen noise strengths and types especially if the amount of noise the model is tested on is lower compared to the one present in the training data.

\begin{table}[t!]
	\centering
	\begin{center}
		\begin{tabular}{ l | c | c | c | c }\hline
			 & Precision & Recall & AP & F1 \\ \hline \hline% \hline
			\tab{(a)} \tab{Model trained on data with strong}&  &  &  &  \\ 
			\tab{} \tab{Gaussian noise and tested on data with:} &  &  &  &  \\ 
			\qquad\quad w/o noise & 91.2 & 91.4 & 82.1 & 91.3 \\ 
			\qquad\quad weak Gaussian noise & 92.3 & 89.3 & 83.3 & 90.8 \\ 
			\qquad\quad moderate Gaussian noise & 91.0 & 87.7 & 83.3 & 89.3 \\ 
			\qquad\quad strong Gaussian noise* & 84.7 & 78.2 & 78.5 & 81.3 \\
			\qquad\quad weak stripes & 87.6 & 81.3 & 71.9 & 84.3 \\
			\qquad\quad moderate stripes & 65.9 & 58.4 & 51.0 & 61.9 \\ 
			\qquad\quad strong stripes & 46.4 & 44.8 & 37.8 & 45.6 \\\hline
			
			 \tab{(b)} \tab{Model trained on data with weak stripes} &  &  &  &  \\ 
			 \tab{} \tab{and tested on data with:} &  &  &  &  \\ 
			 \qquad\quad w/o noise \qquad\qquad\qquad\qquad\qquad\quad\ & 91.6 & ~~91.6~~ & ~~71.1~~ & ~~91.6~~ \\
			 \qquad\quad weak Gaussian noise & 90.1 & 82.8 & 86.6 & 86.3 \\
			 \qquad\quad moderate Gaussian noise & 83.1 & 76.1 & 85.8 & 79.5 \\
			\qquad\quad strong Gaussian noise & 61.6 & 41.0 & 64.2 & 49.3 \\
			\qquad\quad weak stripes* & 90.9 & 90.5 & 88.2 & 90.7 \\ 
			\qquad\quad moderate stripes & 83.4 & 77.1 & 79.4 & 80.1 \\
			\qquad\quad strong stripes & 61.7 & 53.1 & 58.3 & 57.1 \\\hline
		\end{tabular}
	\end{center}
	\caption{Detector performance metrics (precision, recall, average precision (AP), and maximum F1 score) for a model trained on images with (a) strong Gaussian noise $(\sigma = 0.5)$ and (b) weak stripe noise $(A=0.1)$. Each of the two models is tested on all data sets containing a different strength and/or source of noise. The type of noise marked with a star represents the corresponding data set on which the model was trained. The Gaussian noise is added to the normalized BEC density with mean zero and standard deviations $\sigma=0.1$ (weak), $\sigma=0.2$ (moderate) $\sigma=0.5$ (strong). The stripe pattern resulted from adding a sinusoidal modulation to the normalized density with amplitudes $A = 0.2$ (weak), $A = 0.5$ (moderate), $A = 1.0$ (strong).}
	\label{tab:03}
\end{table}

\section{Generalization to different trap geometry}\label{ap:ring}

\begin{figure}[t]
	\centering
	\hspace*{-32mm}
	\begin{subfigure}[b]{0.3\textwidth}
		\centering
		\includegraphics[width=\textwidth]{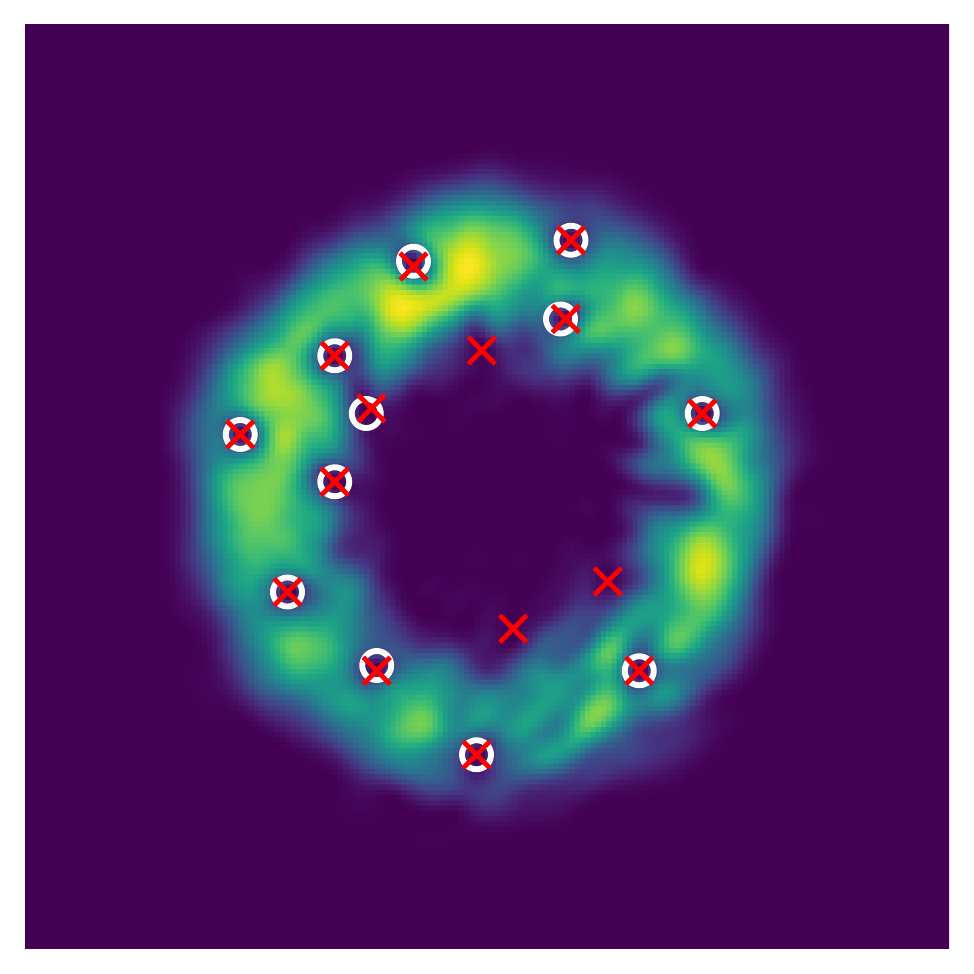} 
		\caption{}
	\end{subfigure}
		\hspace*{1mm}
	\begin{subfigure}[b]{0.3\textwidth}
		\centering
		\includegraphics[width=\textwidth]{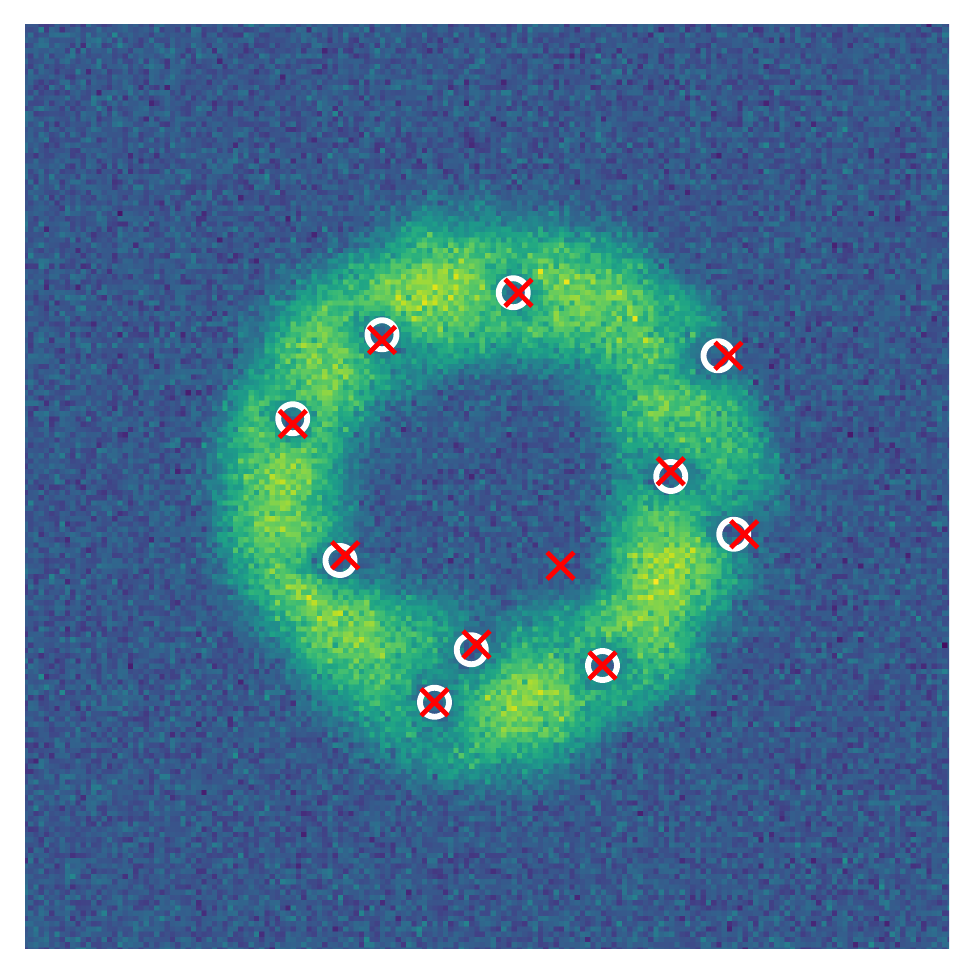} 
		\caption{}
	\end{subfigure}
		\hspace*{1mm}
	\begin{subfigure}[b]{0.3\textwidth}
		\centering
		\includegraphics[width=\textwidth]{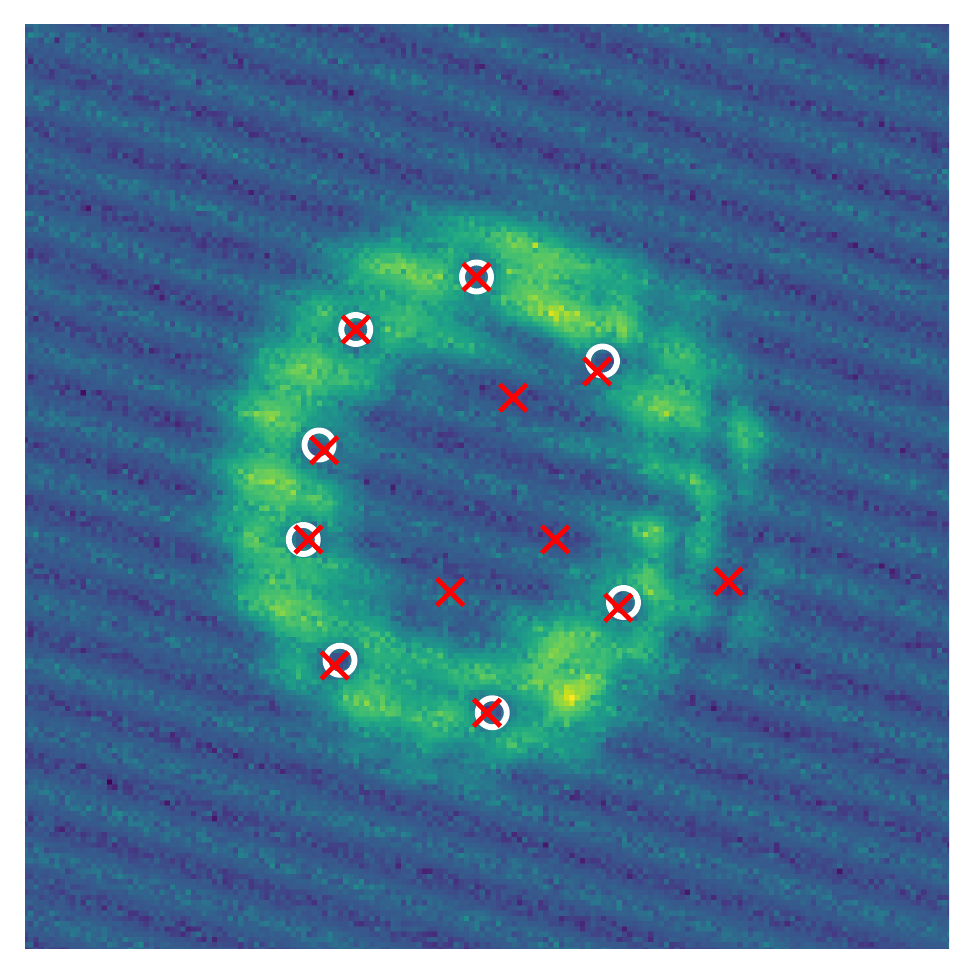} 
		\caption{}
	\end{subfigure}
	\hspace*{-37mm}\hfill\\
	
	\hspace*{-32mm}
	\begin{subfigure}[b]{0.3\textwidth}
		\centering
		\includegraphics[width=\textwidth]{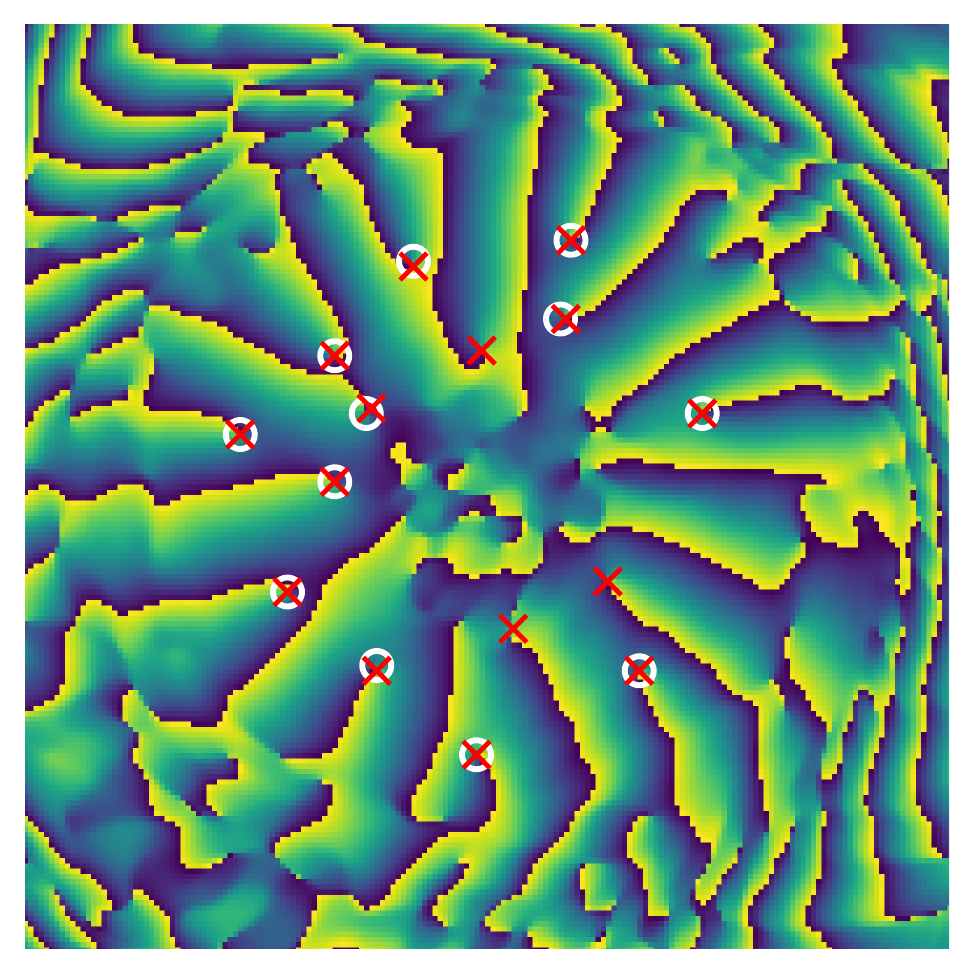}
		\caption{}
	\end{subfigure}
	\hspace*{1mm}
	\begin{subfigure}[b]{0.3\textwidth}
		\centering
		\includegraphics[width=\textwidth]{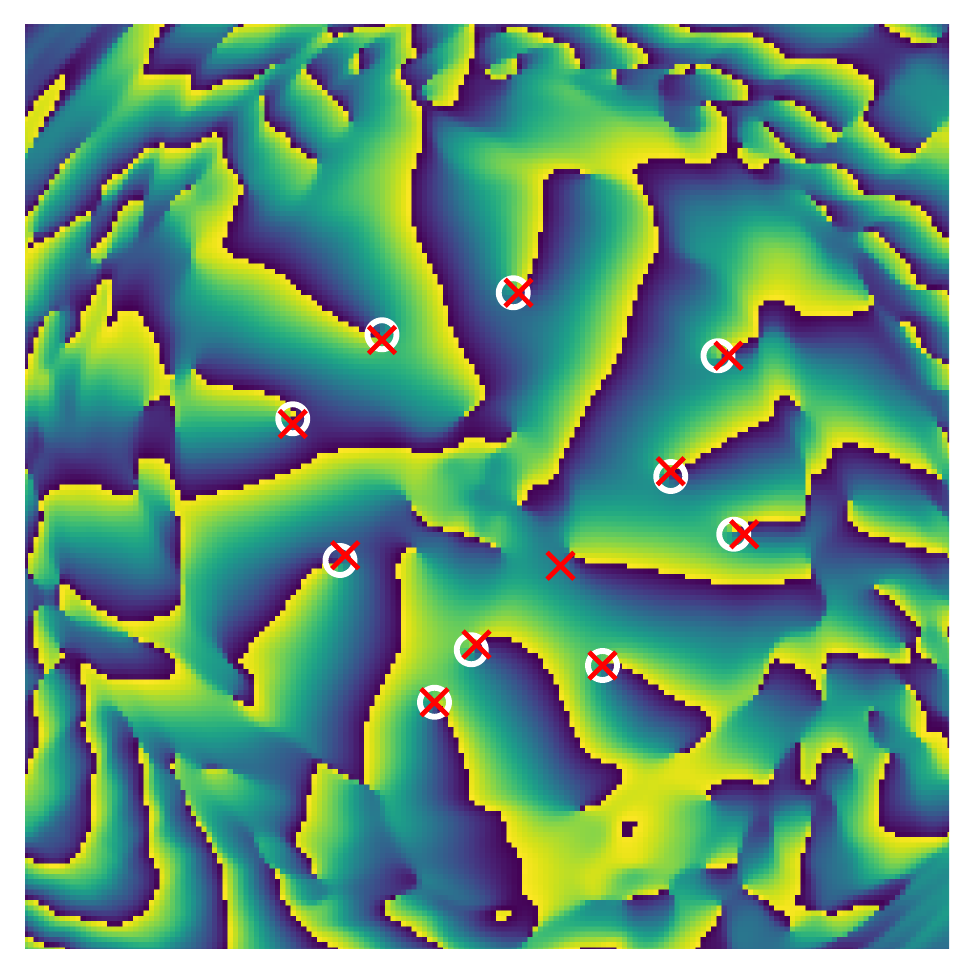}
		\caption{}
	\end{subfigure}
	\hspace*{1mm}
	\begin{subfigure}[b]{0.3\textwidth}
		\centering
		\includegraphics[width=\textwidth]{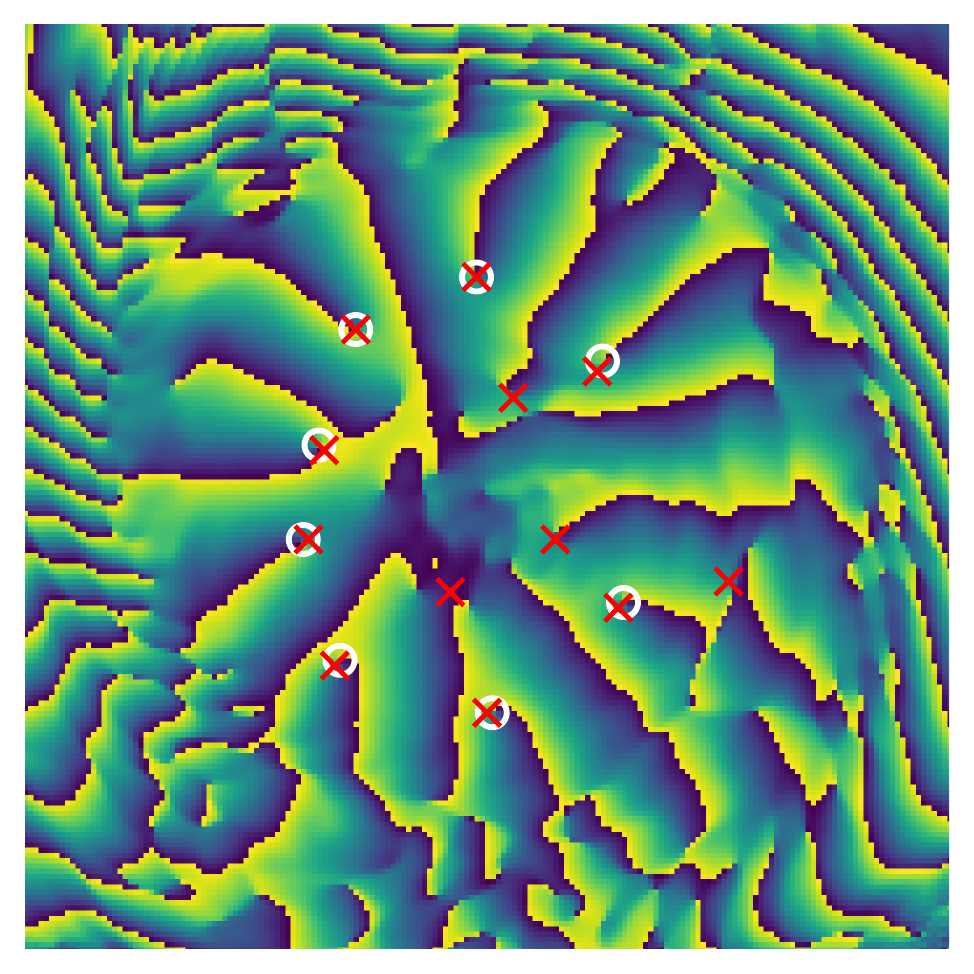}
		\caption{}
	\end{subfigure}
	\hspace*{-37mm}
	\caption{Density (a)-(c) and corresponding phase (d)-(f) configurations for a BEC in a ring shaped trap. A weak Gaussian noise is added to the density in (b) and a stripe pattern is added to the density in (c). Each density image is fed through the corresponding trained network from the main text with the resulting predictions displayed as red crosses and the ground truth as white circles. The phase profile is shown for pure visualization purposes that is, only the density images were used as input to the CNN.} 
	\label{fig:08}
\end{figure}

The networks considered in this paper have been trained solely on BEC configurations in a harmonic trap. To examine whether the same model can be used to detect vortices in different trap geometries, we generate additional images of a BEC in a ring shaped trap. The potential in the GPE (equation \eqref{eq:01}) is replaced by $V=\frac{1}{2} m \omega_{r}^{2}(r-r_0)^2$ with $r_0$ being the toroidal radius and $\omega_r$ the radial trapping frequency. An example of a resulting condensate density and phase profile after real time evolution is shown in figures \ref{fig:08}(a),(d). We test some of our previously trained networks on these new configurations without any further training and indicate the predictions together with their ground truth in figure \ref{fig:08}. The model is able to accurately locate the vortices within the high density regions for images without (figure \ref{fig:08}(a)) and with noise (figure \ref{fig:08}(b)-(c)) while it detects additional vortices at the border of the condensate.  Increasing the confidence threshold will likely result in fewer false positive detections. Due to the overall good performance we observe for the ring shaped potential, we expect that our trained models generalize well to other trap geometries of similar symmetry.

\newpage
\section*{References}

\bibliographystyle{iopart-num.bst}

\bibliography{references.bib}

\end{document}